# Ultrafast All-Optical Polarization Control in a Plasmonic Metasurface by Selective Hot-Electron Heating of Orthogonal Gold Nanorods

*Nikolaos Matthaiakakis,[1,2]* Léa Testé[2], George Kakarantzas[2]*

[1] Institute of Nanoscience and Nanotechnology, National Center for Scientific Research "Demokritos", 15341 Ag. Paraskevi, Greece

[2] *Theoretical and Physical Chemistry Institute, National Hellenic Research Foundation, NHRF, 48 Vassileos Constantinou Ave., 11635 Athens, Greece*

**Abstract:** Metamaterials tailor light's degrees of freedom through static designs, but sub-picosecond, all-optical control without sacrificing signal throughput remains challenging. Here, a metasurface is presented whose unit cell contains two independent, orthogonal and spectrally detuned Au nanorod resonators. A femtosecond pump polarized along either nanorod selectively heats its electrons, modulating one polarization channel while the other serves as an essentially unperturbed phase reference, turning a static wave retarder into an ultrafast, all-optically reconfigurable one. The transient response is computed by coupling a Three-Temperature Model, benchmarked against published experiments, with Finite-Difference Time-Domain simulations. Hot electrons damp and red-shift the pumped resonance, modulating the amplitude ratio and phase retardation of the two channels. In transmission at normal incidence and sub-mJ cm$^{-2}$ fluence, a 45° probe undergoes changes of ≈10° in ellipticity and ≈15° in optical rotation within ≈100 fs (relaxation time ≈4 ps), while its transmittance (≈0.4) changes by at most ≈0.05. A second, delayed pump along the other nanorod raises these changes to ≈15° and ≈22° and, for a suitably polarized probe, cancels the intensity change of the first within ≈60 fs while the polarization stays changed. Selective all-optical control of independent resonators offers a practical route to transmissive ultrafast polarization control.

## Introduction

The increasing demand for higher data rates, multifunctionality, and low latency in modern communication, optical computing, and sensing systems necessitates the development of advanced solutions to manipulate light's degrees of freedom, such as amplitude, phase, polarization, and angular momentum, on ultrafast timescales. Consequently, artificial optical metamaterials have emerged as a leading platform for advanced light-matter interactions[1–6].

Complex control of the degrees of freedom of light, including polarization conversion and arbitrary wavefront shaping, is typically achieved by superimposing distinct optical modes and exploiting their amplitude and phase interference [7–9]. The simultaneous excitation of these modes (mode-mixing) can be intrinsic, dictated by static geometrical designs [10–12], or extrinsic, achieved by dynamically breaking the excitation or detection symmetry[7,8,13,14]. However, relying on a single static metamaterial for dynamic control is inherently limited; different functionalities require varying amplitude-phase relations that remain inextricably tied to the fixed state of the meta-atoms.

To overcome these limitations, dynamically tunable metamaterials driven by external stimuli have been proposed [3,8,13,15–17]. All-optical ultrafast tuning, achieved via high-power femtosecond laser pulses, is a highly promising strategy [9,18–30]. It bypasses the resistance–capacitance (RC) time constant limitations of electrical gating, enabling sub-picosecond response times, massive operational bandwidths, and seamless integration with photonic systems. Beyond the fundamental exploration of nonlinear light-matter interactions, mastering the temporal evolution of the instantaneous polarization state on such timescales is practically critical for next-generation optical technologies [18,31,32]. In modern telecommunications and hyperscale data centres, ultrafast polarization control is essential for expanding bandwidth capacities through multiplexing and dynamic, sub-nanosecond optical routing, minimizing the thermal and latency bottlenecks inherent to electronic processors. Furthermore, in the rapidly accelerating fields of optical computing and quantum information processing, the ability to rapidly and coherently switch polarization states is critical. It provides the necessary framework for executing parallel all-optical logic operations[33,34], performing analog real-time image processing[31,35,36], and establishing secure quantum key distribution channels where fragile quantum data must be encoded and manipulated non-destructively[37,38]. Metasurfaces are increasingly explored as compact elements for manipulating and measuring such polarization-encoded quantum states[39,40].

Achieving ultrafast, all-optical control over light polarization, specifically the dynamic modulation of ellipticity and optical rotation, remains fundamentally constrained by the physical trade-offs between modulation depth, recovery speed, and absolute transmission efficiency. Platforms relying on free-carrier recombination or structural phase transitions face inherent bottlenecks in recovery speed; for instance, hybrid silicon metasurfaces[41] are limited by relaxation times of hundreds of picoseconds. Conversely, while materials operating in the epsilon-near-zero (ENZ) regime facilitate massive, ultrafast nonlinear modulations [42], they traditionally suffer from severe transmissive attenuation. Although reflective perfect absorbers can circumvent these losses[43], their completely opaque architectures render high-efficiency transmissive control elusive. To bridge this gap, recent efforts have shifted toward active plasmonic and dielectric metasurfaces driven by hot-electron transfer [44] and photogenerated carrier nonlinearities [32]. However, these promising architectures remain practically constrained: high-contrast dielectric platforms rely on strictly reflective geometries that prevent transmissive routing, while transmissive plasmonic devices have either produced small transient anisotropies, such as a dichroism of a few percent in $\Delta T/T$[19,45], or reached polarization rotations of tens of degrees only at oblique incidence and control fluences of ≈5–10 mJ $cm^{-2}$, starting from a low transmission of the *p*-polarized component (≈3.6%) and with strong intensity modulation[46]. Consequently, to achieve macroscopic transmissive polarization shifts, state-of-the-art designs frequently must resort to low-efficiency frequency conversion processes, such as second-harmonic generation[47], or use hyperbolic metamaterials at oblique incidence with a strongly reduced output amplitude[48], drastically reducing absolute signal throughput.

Furthermore, most of these all-optical switches rely on uniform spatial pumping that preserves the initial mode-symmetry of the metasurface, precluding the selective mode-mixing required for complex polarization control. Consequently, a critical need remains for a transmissive platform capable of inducing pronounced transient anisotropy to deliver significant, sub-picosecond polarization shifts without sacrificing absolute transmission efficiency.

In most state-of-the-art devices, a structurally homogeneous metasurface is pumped, yielding a uniform spectral shift of all supported modes, preserving the initial mode-symmetry and precluding selective mode-mixing that is needed for selective control over different degrees of freedom of light. Achieving selective control of individual modes supported by the metasurface would increase the control freedom of these devices. To tackle this challenge, a metasurface design that combines extrinsic mode-mixing with the ultrafast nonlinear optical response of noble metals is selected. Specifically, an orthogonal nanorod dimer meta-atom architecture that separates orthogonal localized surface plasmon (LSP) modes into two independently excitable Au nanorods is employed for this purpose. Orthogonal, spectrally detuned dipolar resonances impose a phase retardation between orthogonal field components, as in reflective plasmonic wave retarders, where a single nanocross or nanobrick supports both dipoles and the retardation is fixed by its geometry[49]. Here, the two resonances are instead provided by two independent, orthogonally oriented resonators, each of which can be tuned individually: statically through its own geometry and dynamically by a pump polarized along it[26]. In this way, the static wave retarder becomes an all-optically reconfigurable one that operates in transmission on ultrafast timescales. In the visible-to-near-infrared regime, gold's nonlinearity is driven by extreme, non-equilibrium electron heating following femtosecond laser absorption. This process triggers the instantaneous generation of non-thermal electrons, which rapidly thermalize via electron-electron scattering into a hot electron gas, before ultimately cooling through electron-phonon coupling to the lattice. The resulting smearing of the Fermi-Dirac distribution drastically alters interband transitions and modifies intraband Drude damping. By matching the pump polarization to the orientation of one specific Au nanorod, that element can be selectively excited while the orthogonal nanorod remains relatively unperturbed. This transient asymmetric perturbation alters the relative amplitude and phase between the orthogonal LSPs, providing on-demand, ultrafast control over the amplitude and polarization of the transmitted light.

Polarization-selective excitation of plasmonic metasurfaces has previously been exploited to induce a broadband transient dichroism in $C_4$-symmetric gold nanocrosses, isotropic at equilibrium, through the anisotropic distribution of hot carriers within each nanocross[45]. This transient symmetry breaking has since been used to reconfigure ultrafast dichroism and to switch quasi-bound states in the continuum[19,25]. In these nanocrosses, the induced anisotropy is a transient dichroism of ≈2% in $\Delta T/T$ that vanishes within a few hundred femtoseconds, as electron diffusion restores a homogeneous hot-carrier distribution. In the present design, by contrast, either of the two independent resonators can be heated selectively, and both can be heated sequentially by two delayed pumps. The hot electrons remain confined to the pumped resonator, so that the full change of its permittivity acts on one channel while the other remains essentially unperturbed. Because the two resonators are electrically isolated, this anisotropy is not erased by diffusion and persists until the electrons cool through electron–phonon coupling, and the static design determines how it maps onto the azimuth and ellipticity of the transmitted light.

To accurately capture and validate this transient response, a time-resolved Three-Temperature Model (3TM)[20,50] is combined with full-wave finite-difference time-domain (FDTD) simulations. The transient response of the Au metasurface pumped by an 800 nm femtosecond optical pulse is investigated. By tracking the non-equilibrium dynamics of the non-thermal energy density ($N$), the thermal electron temperature ($\Theta_e$), and the lattice temperature ($\Theta_l$), the temporal evolution of the complex permittivity ($\Delta\varepsilon$) across a broad spectrum can be evaluated explicitly accounting for both Drude intraband variations and the X- and L-point interband transitions.

The simulations reveal that an intense non-thermal electron population, followed by strong electronic heating, induces a rapid transient permittivity change that dynamically modulates the optical anisotropy of the unit cell. This asymmetric perturbation yields a significant, time-dependent modulation of the polarization states within the 700–850 nm wavelength range. Predicting, for a 45° probe, shifts of ≈10° in ellipticity (η) and ≈15° in optical rotation (α) that rise within ≈100 fs, with an absolute transmission value of 40% (changing by at most ≈5 percentage points under pumping), this framework goes beyond static structural tuning and eases the common trade-off between modulation depth and insertion loss. A second pump, polarized along the other nanorod and delayed by 500 fs, modulates the other channel and raises these shifts to ≈15° in ellipticity and ≈22° in optical rotation. For a suitably polarized probe, it also cancels the intensity change caused by the first pump within ≈60 fs, while the polarization state is displaced by ≈30° on the Poincaré sphere, so that the duration of the intensity modulation is set by the pump delay rather than by the picosecond relaxation of the hot electrons. The probe polarization, in turn, selects whether the pump mainly modulates the intensity, the ellipticity or the azimuth of the transmitted light, with larger changes at lower transmission (e.g. a rotation of ≈25° for a probe at 27°). Simulations of the substrate stack, the probe duration and a rescaled design on bulk substrates further show that this modulation remains accessible with standard femtosecond sources at normal incidence, establishing the selective all-optical control of independent resonators as a practical route to transmissive ultrafast polarization control.

**Proposed design and principle of operation**

The metasurface consists of Au orthogonal nanorod dimers on a low-index dielectric substrate (n = 1.2, e.g. MCM-41 (Mobil Composition of Matter No. 41) nanoporous silica) to enhance near-field confinement and suppress parasitic Kerr nonlinearities. For experimental transmissive measurements, this porous film is deposited onto a transparent, high-thermal-conductivity backing wafer (e.g., sapphire or $CaF_2$). Because the porous layer introduces thermal resistance, it should be kept ultrathin (120–250 nm) to facilitate rapid heat dissipation while fully encapsulating the plasmonic near-field (suitable, e.g., for 2 kHz repetition rates, typical for such experiments). Alternatively, for high-repetition-rate experiments where the thermal barrier is prohibitive, the Au dimers can be fabricated directly on bulk fused silica or $CaF_2$. To compensate for this higher refractive index environment, optimal resonant excitation is achieved either by tuning the pump wavelength to match the redshifted LSP resonance, or by reducing the nanorod lengths by ≈18–24% and the Au thickness to 50 nm to maintain the 800 nm resonance (see Supporting Information, Section S5).

As illustrated in **Figure 1a**, the unit cell features a 500 nm periodicity along both the x- and y-axes. The Au nanorods have a uniform thickness of 100 nm. To establish the required geometric anisotropy, the first antenna (60 nm wide and 170 nm long) is aligned parallel to the y-axis and displaced by +100 nm from the unit cell centre along that same axis. Conversely, the second antenna (210 nm long and 60 nm wide) is aligned parallel to the x-axis and offset by −100 nm from the centre along the y-axis.

Optically, the transmission of an incident linearly polarized wave through this structure is described by the scattering matrix:

$$\begin{bmatrix} E_{x,tr} \\ E_{y,tr} \end{bmatrix} = \begin{bmatrix} \tau_{xx} & \tau_{xy} \\ \tau_{yx} & \tau_{yy} \end{bmatrix} \begin{bmatrix} E_{x,inc} \\ E_{y,inc} \end{bmatrix} \quad (1)$$

Both nanorods are centred on the plane $x = 0$, which is a mirror plane of the unit cell. At normal incidence, x- and y-polarized fields have opposite parity under this mirror operation and cannot be converted into each other. Hence $\tau_{xy}$ = $\tau_{yx}$ = 0, also under x- or y-polarized pumping, which preserves this symmetry. The response is thus described by $\tau_{xx}$ and $\tau_{yy}$, each dominated near resonance by the longitudinal dipole mode of one nanorod, so that each nanorod acts as an independent polarization channel (**Figure 1b**). Specifically, the x-oriented antenna supports a longitudinal dipole mode ($LSP_x$), excited by x-polarized light, with a static transmission dip at ≈800 nm (**Figure 1c**). Similarly, the y-oriented antenna supports a longitudinal dipole mode ($LSP_y$), excited only by y-polarized light, with a static resonance at ≈740 nm. The spectral evolution of the transmission amplitudes, $|\tau_{xx}|$ and $|\tau_{yy}|$, is accompanied by a pronounced inherent phase retardation, $\Delta\mathrm{Arg}(\tau)$, between the two channels (**Figure 1d**).

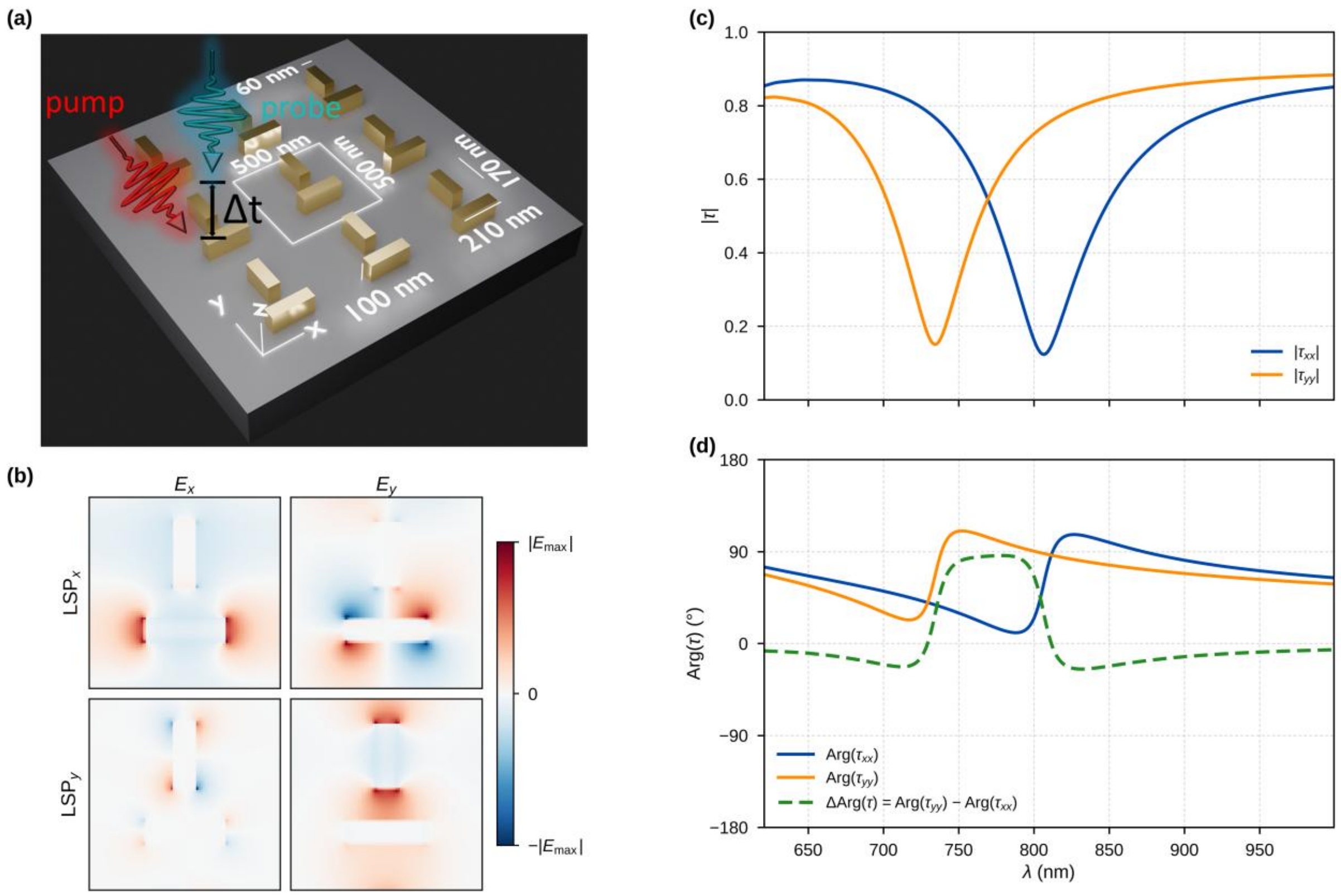

**Figure 1.** Static linear optical response and resonant near fields of the orthogonal nanorod dimer metasurface. **(a)** Schematic illustration of unit cell geometry. **(b)** Instantaneous near-field snapshots of the total field, $E_x$ and $E_y$ (each normalized to its maximum), in the mid-plane of the nanorods, computed by FDTD for normally incident plane-wave pulses that excite the $LSP_x$ resonance (x-polarized, centred at 810 nm; top) or the $LSP_y$ resonance (y-polarized, centred at 740 nm; bottom). The field component parallel to the excited nanorod shows its longitudinal dipole mode; the four-lobed patterns of the orthogonal component are odd under $x \to -x$ and do not contribute to cross-polarized transmission. **(c)** Simulated amplitude spectra of the transmission coefficients, $|\tau_{xx}|$ (blue solid line) and $|\tau_{yy}|$ (orange solid line). The pronounced transmission dips correspond to the static dipole resonances at ≈800 nm and ≈740 nm, respectively. **(d)** Corresponding spectral phase, $\mathrm{Arg}(\tau_{xx})$ and $\mathrm{Arg}(\tau_{yy})$, alongside the relative phase difference ΔArg(τ) (dashed green line), which dictates the inherent phase retardation between the two polarization channels prior to ultrafast optical excitation.

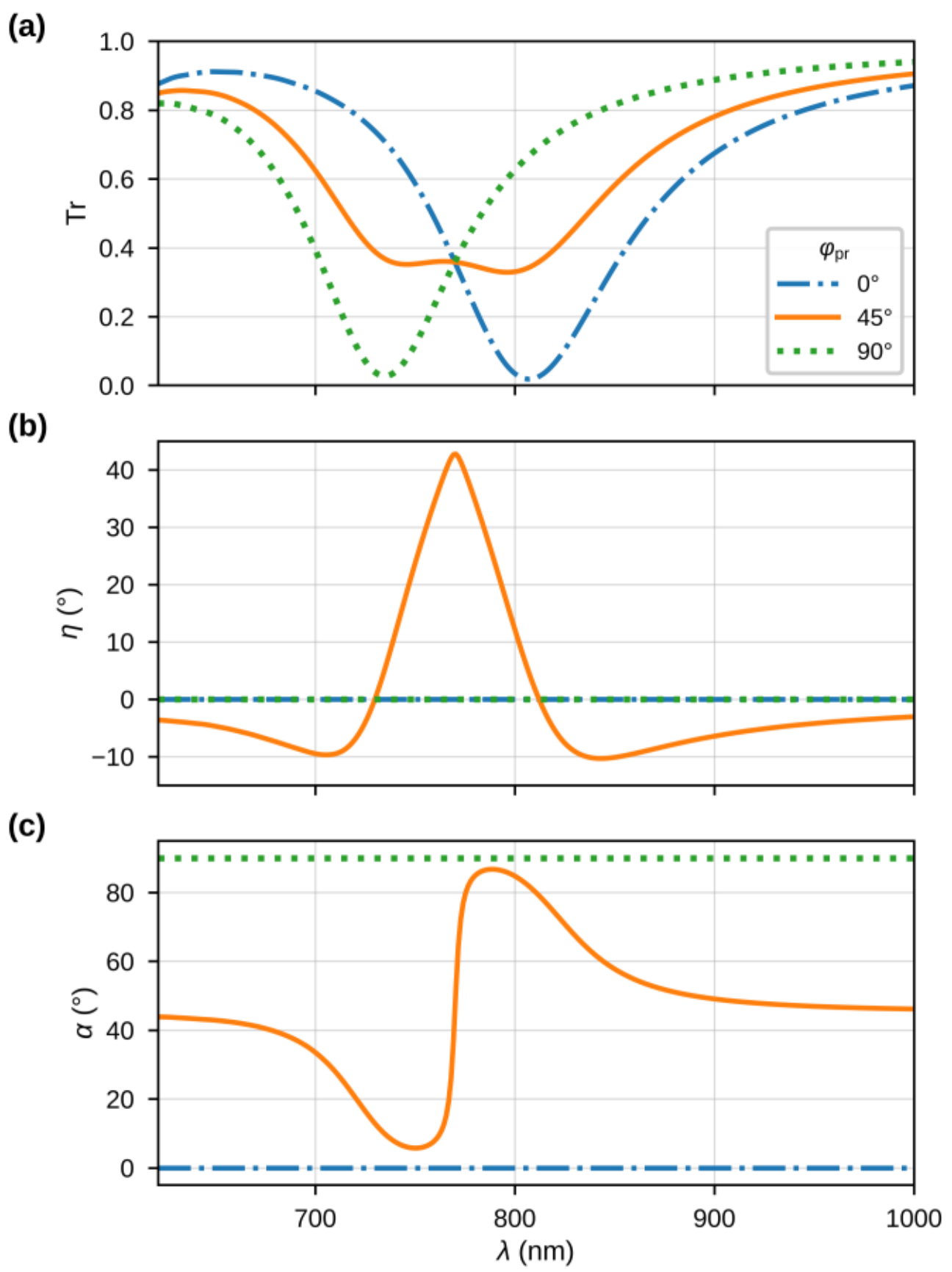

**Figure 2. Static transmission and polarization response of the metasurface as a function of the incident probe azimuth angle. (a)** Simulated transmission spectra (Tr) under linearly polarized excitation at azimuth angles of $\varphi_{\mathrm{pr}} = 0°$ (blue dash-dotted line), $\varphi_{\mathrm{pr}} = 45°$ (orange solid line), and $\varphi_{\mathrm{pr}} = 90°$ (green dotted line). The 0° and 90° spectra isolate the individual orthogonal resonances, while the 45° spectrum reflects their simultaneous excitation. **(b)** Corresponding ellipticity ($\eta$) of the transmitted light. For excitation purely along the structural axes (0° and 90°), the polarization remains strictly linear ($\eta = 0°$). For a probe polarized at 45°, the interference between the orthogonal modes yields a strongly wavelength-dependent ellipticity. **(c)** Optical rotation angle ($\alpha$) of the transmitted light. Similar to the ellipticity, significant optical rotation is only observed for the 45° excitation, demonstrating the static birefringent properties of the dual-nanorod unit cell prior to ultrafast optical pumping.

Ultimately, the complex interference between these direct transmission coefficients governs the final polarization state of the transmitted light. To exploit this, the metasurface is illuminated with a linearly polarized probe pulse at an azimuth angle of $\varphi_{pr} = 45°$, simultaneously exciting both orthogonal modes (**Figure 2a**). While probing purely along the structural axes (0° and 90°) preserves a strictly linear polarization state, the simultaneous 45° excitation leverages the static amplitude and phase differences to induce a strongly wavelength-dependent ellipticity ($\eta$) and optical rotation ($\alpha$), as shown in **Figures 2b** and **2c**, respectively. This static configuration establishes the critical baseline amplitude-phase relationship required to achieve active, ultrafast polarization control upon laser pumping.

**All-Optical Ultrafast Polarization Control**

To achieve ultrafast dynamic control, transient changes are introduced in the dielectric permittivity of gold via a femtosecond pump pulse. To accurately capture these sub-picosecond dynamics, the temporal evolution of the system is governed by a 3TM [20]. The model reproduces the main features of the transient extinction map measured for colloidal gold nanoparticles[51], yields permittivities under strong electronic excitation consistent with previously reported values[52], and reproduces the hierarchy of non-thermal, thermalized-electron and lattice contributions reported for gold nanocross metasurfaces[45] (Supporting Information, Section S3, Figures S2–S4). This model tracks the energy density of the non-thermal electrons ($N$), the temperature of the thermalized electron gas ($\Theta_e$), and the lattice temperature ($\Theta_l$):

$$\frac{dN(t)}{dt} = -\alpha_s N(t) - bN(t) + P_{abs}(t), \qquad (2)$$

$$C_e \frac{d\Theta_e(t)}{dt} = -G\big(\Theta_e(t) - \Theta_l(t)\big) + \alpha_s N(t), \qquad (3)$$

$$C_l \frac{d\Theta_l(t)}{dt} = G\big(\Theta_e(t) - \Theta_l(t)\big) + bN(t), \qquad (4)$$

where $P_{abs}$ is the absorbed power density from the pump pulse, $\alpha_s$ and $b$ are the scattering rates defining non-thermal energy transfer to the electron gas and lattice respectively, $C_e$ and $C_l$ are the volumetric heat capacities, and $G$ is the electron-phonon coupling constant. This multi-stage energy transfer drastically modifies both the intraband Drude damping and the interband transitions, yielding a total transient complex dielectric permittivity, $\Delta\varepsilon(\omega, t) = \Delta\varepsilon_{\text{intra}}(\omega, t) + \Delta\varepsilon_{\text{inter}}(\omega, t)$. To trace this transient optical response, the 3TM dynamics are included in the FDTD simulations. Detailed information on the 3TM and the permittivity calculations is given in the Supporting Information.

A 70 fs pump pulse at λ = 800 nm with a fluence of $F \approx 840$ μJ/cm$^2$ is used to excite the metasurface. By polarizing the pump exclusively along the x-axis ($\varphi_{\text{pu}} = 0°$), the pump couples predominantly to the $LSP_x$ mode, confining the absorption almost entirely within the x-oriented nanoantenna (**Figure 3a**). Consequently, this selective excitation drives an instantaneous surge in non-thermal energy density (**Figure 3b**). Through rapid electron-electron thermalization, the thermal electron gas temperature ($\Theta_e$) in the x-oriented antenna exceeds 2000 K within ≈250 fs and peaks at ≈2290 K at ≈0.7 ps (Figure 3c), followed by a slower picosecond energy transfer to the lattice ($\Theta_l$) via electron-phonon coupling (**Figure 3d**). Crucially, throughout this intense thermodynamic cycle, the orthogonal y-oriented antenna remains largely unperturbed (it receives ≈1% of the absorbed pump power; Supporting Information, Section S8), so that the optical anisotropy of the unit cell is modulated on a sub-picosecond timescale.

The heating of the electron gas modifies the permittivity of the pumped x-oriented nanoantenna (**Figure 3e**): the non-thermal electrons and the smearing of the Fermi–Dirac distribution alter the X- and L-point interband transitions, and the increased electron–electron and electron–phonon scattering raises the Drude damping. Around 800 nm, both Re($\Delta\varepsilon$) and Im($\Delta\varepsilon$) are positive. The increase of Im($\varepsilon$) damps and broadens $LSP_x$, while the less negative Re($\varepsilon$) red-shifts it slightly, because the dipolar resonance condition Re($\varepsilon_{Au}$) ≈ $-\chi\varepsilon_d$ ($\chi$: shape factor; $\varepsilon_d$: permittivity of the surroundings) is then met at a longer wavelength. Accordingly, the differential transmission (**Figure 3i**) shows a bleaching of the $LSP_x$ transmission dip (**Figure 3f**) flanked by induced absorption in its wings. The y-oriented nanoantenna receives only ≈1% of the absorbed pump power and is kept at its equilibrium permittivity (Supporting Information, Section S8, estimates the effect of this absorption); the induced absorption near 730 nm stems from the broadened wing of $LSP_x$.

Because the unit cell is already anisotropic at equilibrium (**Figures 1** and **2**), the selective excitation of the x-oriented nanoantenna modulates the degree of this anisotropy. Since $\tau_{xy} = \tau_{yx} = 0$, the field transmitted for a probe polarized at $\varphi_{pr} = 45°$ is proportional to ($\tau_{xx}$, $\tau_{yy}$), and its azimuth $\alpha$ and ellipticity $\eta$ follow from the amplitude ratio $\tan\psi = |\tau_{yy}|/|\tau_{xx}|$ and the retardation $\delta = \Delta\mathrm{Arg}(\tau) = \mathrm{Arg}(\tau_{yy}) - \mathrm{Arg}(\tau_{xx})$[53]:

$$\tan 2\alpha = \tan 2\psi \cos\delta, \qquad \sin 2\eta = \sin 2\psi \sin\delta \qquad (5)$$

The pump modifies essentially only $\tau_{xx}$: near the $LSP_x$ resonance, where Arg($\tau_{xx}$) varies steeply (**Figure 1d**), it changes Arg($\tau_{xx}$) by up to ≈32°, whereas Arg($\tau_{yy}$) changes by less than 1° (**Figure S7**). Through Equation (5), the resulting changes of the amplitude ratio, tan $\psi$ (transient dichroism), and of the retardation $\delta$ (transient birefringence) shift the ellipticity by up to ≈10° and the optical rotation by ≈15°, or by up to ≈20° where the transmitted light is nearly circular (**Figures 3g,h,j,k**). The change appears mainly in $\eta$ where the static retardation is small (≈810–840 nm), and mainly in $\alpha$ where it approaches 90° (≈765–800 nm). The modulation peaks within ≈100 fs of the pump arrival, when the non-thermal energy density $N$ is maximal. It is then sustained by the hot thermalized electrons ($\Theta_e$) and decays over a few picoseconds as they transfer their energy to the lattice.

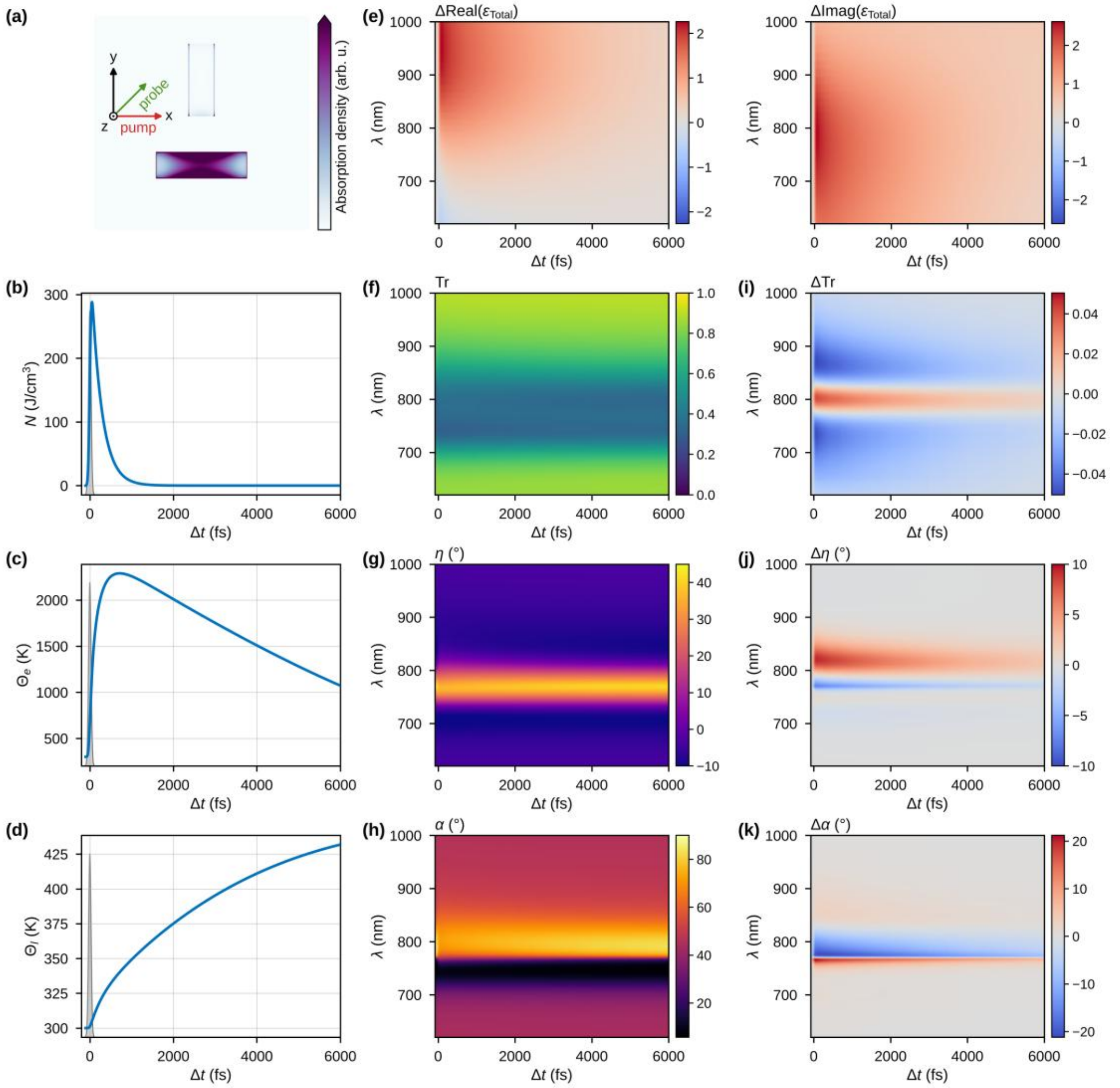

**Figure 3.** **Ultrafast thermodynamic and transient optical response of the orthogonal nanorod dimer metasurface**. **(a)** Simulated spatial absorption density profile, demonstrating the selective, asymmetric excitation of the x-oriented nanoantenna under an x-polarized femtosecond pump pulse. **(b–d)** Temporal evolution of the internal physical dynamics derived from the 3TM for the excited antenna: **(b)** non-thermal electron energy density ($N$), **(c)** thermal electron temperature ($\Theta_e$) peaking at ≈2290 K at ≈0.7 ps, and **(d)** lattice temperature ($\Theta_l$) driven by subsequent electron-phonon thermalization. The grey shaded region at $\Delta t = 0$ fs indicates the incident pump pulse profile. **(e)** Corresponding transient modulations in the real ($\Delta$Real($\varepsilon_{Total}$)) and imaginary ($\Delta$Imag($\varepsilon_{Total}$)) components of the gold dielectric function. **(f, i)** Absolute transmission ($Tr$) and the transient differential transmission ($\Delta Tr$) maps as a function of probe wavelength ($\lambda$) and pump-probe delay ($\Delta t$). **(g, j)** Absolute ellipticity ($\eta$) and its pump-induced transient modulation ($\Delta\eta$), achieving sub-picosecond peak shifts of ≈10°. **(h, k)** Absolute optical rotation ($\alpha$) and the corresponding transient change ($\Delta\alpha$), demonstrating ultrafast dynamic rotation of ≈15°, and up to ≈20° where the transmitted light is nearly circular. The temporal evolutions in panels (i–k) show the ultrafast, active control over light's degrees of freedom via selective structural pumping.

To visualize the evolving polarization state, the normalized Stokes vector of the transmitted light is mapped onto the Poincaré sphere at selected probe wavelengths (**Figure 4a**). At each wavelength, the state moves away from its static position within ≈100 fs, while the non-thermal electron population is maximal, and returns along almost the same path over several picoseconds as the hot electrons transfer their energy to the lattice. The direction and length of this excursion depend on the wavelength through the static retardation (**Figure 1d**). **Figure 4b–e** shows the same dynamics versus delay: the normalized Stokes parameters change by up to ≈0.35, the transmittance by less than 0.04. Overall, the polarization state moves by up to ≈24° on the Poincaré sphere. The largest change of the azimuth (≈20°, **Figure 3k**) occurs near 765 nm, where the transmitted light is nearly circular; where the polarization ellipse is well defined ($|\eta| \leq 20°$), $\Delta\alpha$ reaches ≈15°.

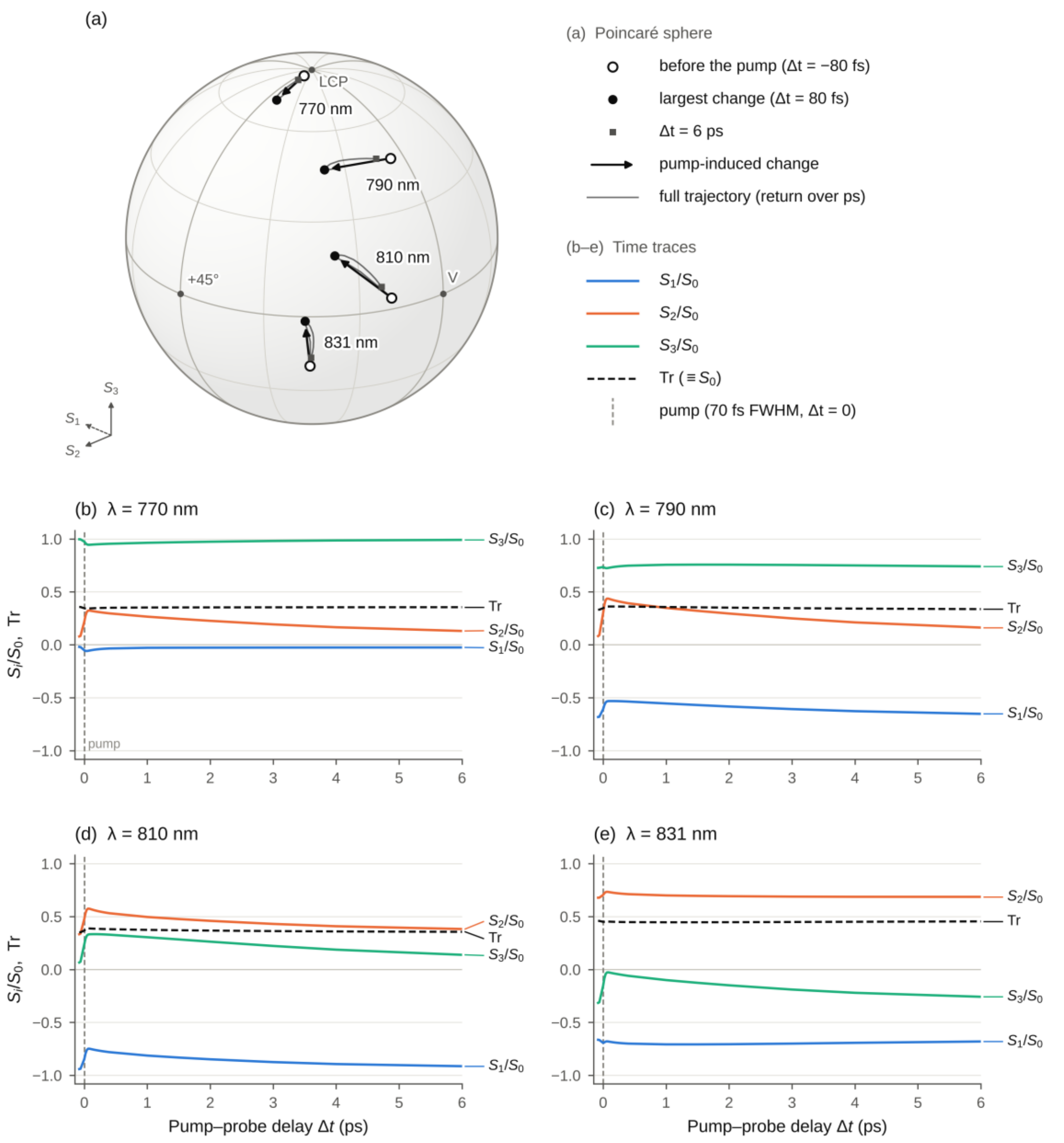

**Figure 4. Dynamics of the transmitted polarization state. (a)** Trajectories of the normalized Stokes vector on the Poincaré sphere at 770, 790, 810 and 831 nm, shown in the same orthographic view. Open circles: before the pump ($\Delta t = -80$ fs); filled circles: largest change ($\Delta t = 80$ fs); squares: $\Delta t = 6$ ps. Arrows show the pump-induced change; thin lines show the full trajectory, including the return over several picoseconds. V: vertical linear; +45°: diagonal linear; LCP: left-handed circular polarization. **(b–e)** $S_1/S_0$, $S_2/S_0$, $S_3/S_0$ and transmittance $\mathrm{Tr} \equiv S_0$ versus $\Delta t$ at the same wavelengths, on a common scale. The dashed line marks the pump pulse ($\Delta t = 0$; 70 fs full width at half maximum, FWHM).

For the x-polarized pump of Figures 3 and 4, the probe polarization provides a further degree of freedom (**Figure 5**). Because the Jones matrix remains diagonal under the pump, a probe linearly polarized at an azimuth $\varphi$ from the x axis is transmitted as $E \propto (\tau_{xx} \cos \varphi, \tau_{yy} \sin \varphi)$, so the simulations with the 45° probe determine the response to any linear probe (Supporting Information, Section S7); direct simulations with the probe at $\varphi = 27°$ agree with this reconstruction to within 0.05° on the Poincaré sphere at every pump–probe delay (Supporting Information, Figure S8). **Figure 5a** shows how far the pump moves the polarization state on the Poincaré sphere, i.e. the angle between the normalized Stokes vectors with and without the pump, and **Figure 5c** the direction of this displacement: 0° for a pure rotation of the polarization ellipse, which moves the state along a parallel of the sphere, and 90° for a pure change of its ellipticity, which moves it along a meridian. Probes polarized along x or y are eigenpolarizations: the pump changes only their intensity, by up to +449% along the pumped nanorod (setting A, at 806 nm: from a transmittance of 0.019 to 0.10; **Figure 5b**) and by less than 3% along the other. At intermediate azimuths, the probe polarization selects which property of the transmitted light is modulated: near $\varphi = 20°$ and 824 nm (setting C), the pump mainly changes the ellipticity ($\Delta\eta = +14.5°$, $\Delta\alpha = -3.4°$; **Figure 5d**), whereas near $\varphi = 27°$ and 802 nm (setting D) it mainly rotates the polarization ellipse ($\Delta\alpha = -24.8°$, $\Delta\eta = +3.3°$; **Figure 5e**); in **Figure 5f**, which shows the longitude $2\alpha$ and latitude $2\eta$ of the state on the sphere, these two changes appear as a vertical (C) and a horizontal (D) step. Rotating the probe towards the pumped axis also increases the largest displacement on the Poincaré sphere, from 24° at $\varphi = 45°$ (setting B) to 39° at $\varphi = 27°$ and 48° at $\varphi = 14°$, at the cost of transmission (0.33, 0.15 and 0.06).

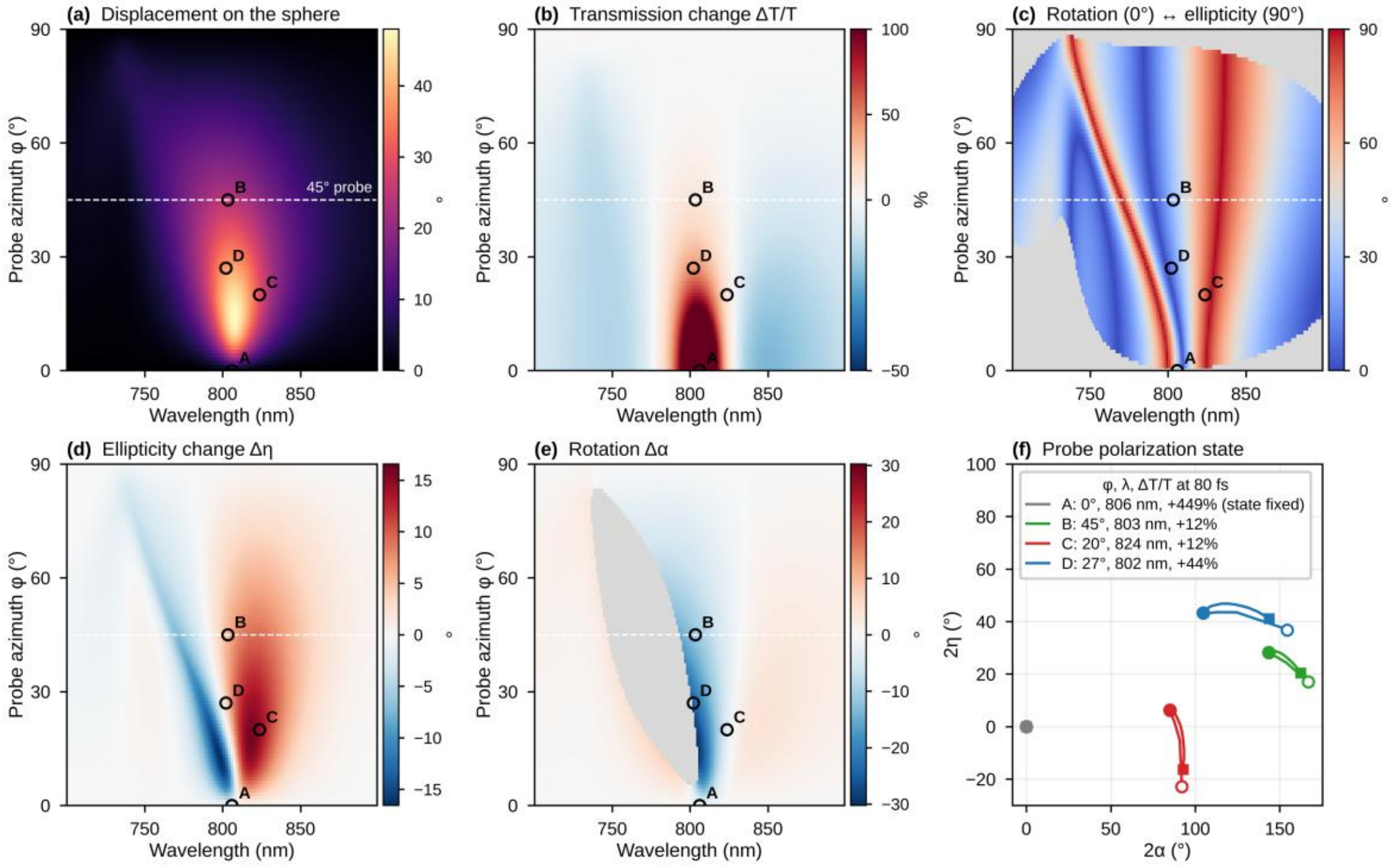

**Figure 5. The probe polarization selects the modulated property.** Pump-induced changes at $\Delta t$ = 80 fs (value at 80 fs minus the value of the unpumped metasurface; x-polarized pump of Figures 3 and 4, ≈840 µJ/cm²) versus probe azimuth $\varphi$ (from the x axis) and probe wavelength, computed from the 45° probe simulations. **(a)** Displacement of the polarization state on the Poincaré sphere, i.e. the angle between the normalized Stokes vectors with and without the pump; a change $\Delta\eta$ of the ellipticity moves the state by $2|\Delta\eta|$ along a meridian, a rotation $\Delta\alpha$ of the ellipse by $\approx 2|\Delta\alpha| \cos 2\eta$ along a parallel. **(b)** Relative change of the transmission $\Delta T/T$ (colour scale clipped at +100%). **(c)** Direction of the displacement relative to the parallel through the unpumped state, $\arctan(|2\Delta\eta|/|2\Delta\alpha \cos 2\eta|)$, with $\eta$ the unpumped ellipticity: 0°, pure rotation of the polarization ellipse (along the parallel, at constant $\eta$); 90°, pure change of ellipticity (along the meridian, at constant $\alpha$); grey: displacement below 2°. **(d)** $\Delta\eta$. **(e)** $\Delta\alpha$; grey: unpumped $|\eta| >$ 20°, where the azimuth is ill-defined. The dashed lines mark the 45° probe of Figures 3 and 4. **(f)** Polarization state of the transmitted probe for the settings A–D marked in (a–e) (○ unpumped, ● 80 fs, ■ 6 ps), in the longitude ($2\alpha$) and latitude ($2\eta$) of the Poincaré sphere, where a rotation of the ellipse appears as a horizontal and a change of ellipticity as a vertical step.

Because either nanorod can be heated selectively, the two polarization channels can also be modulated one after the other. In **Figure 6**, an x-polarized pump at 800 nm ($\Delta t$ = 0) is followed 500 fs later by a y-polarized pump at 740 nm, tuned to $LSP_y$; both have a fluence of ≈840 µJ/cm². The second pump deposits ≈98% of its absorbed power in the y-oriented nanorod (the first pump ≈99% in the x-oriented one) and heats its electrons to ≈2700 K. It changes Arg($\tau_{yy}$) by up to ≈35°, whereas the phase change of $\tau_{xx}$ follows the first pump alone (**Figure 6a,b**); the ≈2% of the second pump absorbed by the x-oriented nanorod, neglected here, would add ≈0.3° (Supporting Information, Section S8). Each pump thus acts mainly in the spectral band of its own resonance (**Figure 6d**; Supporting Information, Section S8): relative to the response to the first pump alone, the second pump moves the polarization state by up to ≈34° on the Poincaré sphere between 720 and 780 nm, near $LSP_y$, but by only a few degrees at and beyond 810 nm. The second pump does not undo the first: a probe wavelength between the two resonances lies on the red side of $LSP_y$ and on the blue side of $LSP_x$, so the red shifts bring $LSP_y$ closer to it but move $LSP_x$ away from it, and the two changes add; only beyond ≈820 nm, on the red side of both resonances, does the second pump move the state back towards its initial value, and by at most ≈2°. At 741 nm (**Figure 6c,d**), for instance, the displacement of the polarization state from its initial state grows from 3° just before the second pump to 36° after it. This displacement, the angle between the normalized Stokes vectors with and without the pumps, comes mainly from a rotation of the polarization ellipse ($\Delta\alpha \approx 20°$, $\Delta\eta \approx 2°$), which moves the state by nearly twice this angle along a parallel of the Poincaré sphere. Over the spectrum, the two pumps change the ellipticity by up to ≈15° (as at 770 nm, **Figure 6d**) and, where the polarization ellipse is well defined ($|\eta| \leq 20°$), the azimuth by up to ≈22° (≈20° at 741 nm, **Figure 6d**).

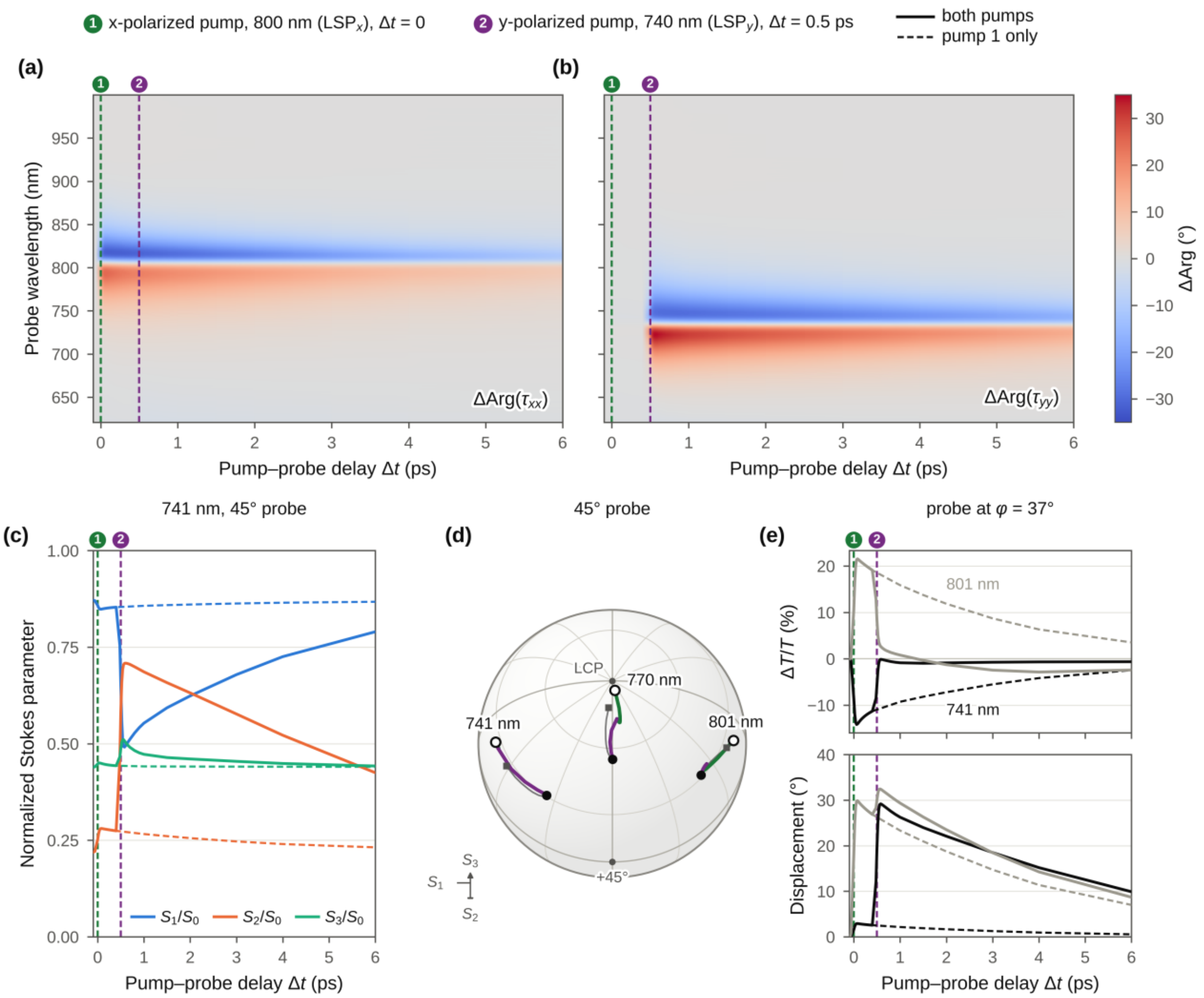

**Figure 6. Sequential, selective excitation of the two nanorods.** An x-polarized pump at 800 nm ($LSP_x$, pump 1) arrives at $\Delta t = 0$ and a y-polarized pump at 740 nm ($LSP_y$, pump 2) at $\Delta t = 0.5$ ps, both at ≈840 μJ/cm² (dashed vertical lines, marked 1 and 2). In (c) and (e), solid lines: both pumps; dashed lines: pump 1 only (no delays between 0.4 and 1 ps). **(a)** $\Delta\mathrm{Arg}(\tau_{xx})$ and **(b)** $\Delta\mathrm{Arg}(\tau_{yy})$ versus probe wavelength and pump–probe delay. **(c)** Normalized Stokes parameters of the transmitted 45° probe at 741 nm. **(d)** Trajectories of the polarization state of the 45° probe on the Poincaré sphere at 741, 770 and 801 nm: green, before pump 2 arrives ($\Delta t \leq 0.4$ ps); purple, after it, up to the largest change; thin line, return up to 6 ps. Open circles: before the pumps; filled circles: largest change; squares: $\Delta t = 6$ ps. **(e)** Relative change of the transmission $\Delta T/T$ (top) and displacement of the polarization state on the Poincaré sphere (bottom) at 741 and 801 nm, for a probe polarized at $\varphi = 37°$ from the x axis (computed from the 45° probe, Supporting Information, Section S7; the solid lines are confirmed by direct simulations at every delay, Figure S8).

The two handles can also be combined. Although the second pump of **Figure 6** does not undo the polarization change of the first, it can undo its change of the transmitted intensity. At either resonance, the pump tuned to it raises the transmittance of its own channel, whose transmission dip is damped, while the transmittance of the other channel falls as the other resonance broadens: at 801 nm, the first pump raises $T_x \propto |\tau_{xx}|^2$ from 0.03 to 0.12 and the second lowers $T_y \propto |\tau_{yy}|^2$ from 0.63 to 0.52; at 741 nm, the first lowers $T_x$ from 0.66 to 0.56 and the second raises $T_y$ from 0.05 to 0.20. A probe polarized at $\varphi = 37°$, which is transmitted with $T_x \cos^2\varphi + T_y \sin^2\varphi$, weights the two channels such that their changes cancel after the second pump, by which time the change of $T_x$ has partly recovered (**Figure 6e**). At 801 nm, its transmission rises by 22% (ΔT/T) after the first pump (12% for the 45° probe) and returns to within ≈3% of its unpumped value within 60 fs of the peak of the second, where it stays up to 6 ps. The remaining deviation, positive at first and negative after ≈1.3 ps, arises because the change of $T_y$ builds up until ≈0.7 ps and then recovers more slowly than that of $T_x$, the y-oriented nanorod being hotter. At 741 nm, a drop of 14% is cancelled to within 1%. With the first pump alone (dashed lines in **Figure 6e**), about half of each change is still present at 2 ps (+12% at 801 nm, −7% at 741 nm). The cancellation does not rely on a fine balance between the two pumps: because $T_x$ and $T_y$ change in opposite directions and the probe azimuth can be chosen freely, a different ratio of their changes would shift the azimuth at which they cancel best rather than prevent the cancellation. With the absorption of each pump in the other nanorod included, this azimuth remains 37° (Supporting Information, Section S8). The polarization state, in contrast, is displaced by 29–32° after the second pump and still by 22–24° at 2 ps. With a single pump, the intensity and polarization changes rise and decay together. The second pump, by contrast, ends the intensity modulation at a time set by the pump delay rather than by the ≈4 ps relaxation, and leaves a polarization modulation at nearly constant transmission.

As summarized in the comparative analysis (Table 1), the current landscape of ultrafast all-optical polarization control is fundamentally defined by the stringent trade-offs between absolute modulation depth, temporal response, and transmissive signal efficiency. While state-of-the-art platforms, such as nonlinear plasmonic metasurfaces or active all-dielectric architectures, report 90° switching of the polarization or a 90° phase retardation, these architectures achieve large modulation at the expense of signal throughput. They rely on either inefficient frequency-conversion processes (e.g., second-harmonic generation) or reflective geometries, both of which complicate cascading in dense nanophotonic networks. Similarly, platforms leveraging ENZ nonlinearities or structural phase transitions typically suffer from strong attenuation in transmission or from recovery extending to the nanosecond scale. Against this backdrop, the proposed orthogonal Au nanorod dimer metasurface offers a favourable operating point. By selectively modulating the optical anisotropy of the unit cell to drive active mode-mixing, this platform goes beyond the static response of conventional linear metasurfaces. Operating directly within the fundamental transmitted beam, it delivers, for a 45° probe, shifts of ≈10° in ellipticity and ≈15° in optical rotation, rising within ≈100 fs, while preserving a ≈40% absolute transmission that changes by at most ≈5 percentage points during dynamic modulation, thereby offering a promising pathway for transmissive all-optical polarization control. A second, delayed pump raises these shifts to ≈15° and ≈22° and, for a suitably polarized probe, ends the intensity modulation within ≈60 fs while the polarization stays modulated.

| Publication (Year) | Platform / Architecture | Spectral Region | Geometry & Efficiency | Polarization Control (Rotation & Ellipticity) | Temporal Response | Driving Mechanism |
|---|---|---|---|---|---|---|
| **Matthaiakakis (2025)**[26] | Dual-Stack Graphene Nanoribbon Metamaterial | Mid-IR (≈12–13 µm) | Transmissive / Reflective. Ultrathin 2D geometry (two graphene nanoribbon layers). | Multiple degrees of freedom: amplitude, phase, polarization and handedness. | Picosecond | Ultrafast selective mode-mixing via pump-induced carrier heating. |
| **Matthaiakakis (2024)**[9] | Nanostructured Graphene Metasurface | Mid-IR (≈10–13 µm) | Reflective. | Linear-to-Circular: pump-driven switching between circular and linear polarization. | Picosecond | All-optical pump-induced free carrier heating. |
| **Wang (2024)** [47] | Nonlinear Plasmonic Metasurface (Au, C3 symmetry) | Near-IR pump to Visible output | Transmissive (SHG). The switched signal is the second harmonic generated from the pump (conversion efficiency not reported). | 90° switching of the second-harmonic polarization (horizontal ↔ vertical; 97% modulation depth). | Hundreds of fs | SHG utilizing broken inversion symmetry. |
| **Crotti (2024)** [32] | All-Dielectric Active Nonlocal Metasurface (AlGaAs) | Near-IR (740–750 nm) | Reflective (on $AlO_x$/GaAs). Static TM reflectance ≈10%, rising to ≈60% under pumping. | Up to 90° (π/2) TM–TE phase shift, turning linear into circular polarization. | Few ps (peak at ≈2 ps) | Band filling effect and photogenerated carriers. |
| **Saha (2023)** [42] | Unpatterned TiN/AZO bilayer (ENZ media) | Vis-NIR | Reflective (50° incidence). Largest modulation near the p-polarized reflectance dips, where ≈80–90% of the light is absorbed. | None (reflectance modulation only) | Tunable (<10 ps to >2 ns, set by the probe wavelength) | Photoexcited free carriers in AZO (fast) and heating of TiN (slow), enhanced near the ENZ wavelengths. |
| **Wang (2021)** [46] | Au nanoantenna array on an ITO (ENZ) film | Near-IR (1.23 µm) | Transmissive (30° incidence). Strong intensity modulation: the transmittance of the p-polarized component rises from ≈3.6% to ≈38% under pumping. | Up to 32.5° rotation of the polarization ellipse. | ≈600 fs | ENZ nonlinearity of ITO coupled to plasmonic nanoantennas (control fluence up to ≈10 mJ/cm²). |
| **Schirato (2020)** [45] | $C_4$-symmetric Au nanocross metasurface | Vis-NIR (probe 565–735 nm; pump 850–950 nm) | Transmissive. Polarization-independent static transmittance. | Transient broadband dichroism (≈2% in $\Delta T/T$). | Sub-ps (peak at ≈100 fs, recovery <1 ps) | Transient optical symmetry breaking by an inhomogeneous, mainly non-thermal hot-carrier distribution. |
| **Cong (2018)** [41] | Hybrid Metal-Silicon Metasurface | THz (0.4–1.0 THz) | Transmissive. Static co- and cross-polarized transmission efficiencies ≈23% and ≈15%. | Output switched from right-handed circular to quasi-linear polarization. | Hundreds of ps (decay constant 667 ps) | Photoexcitation of free carriers in silicon (pump-probe). |
| **Taghinejad (2018)** [44] | Plasmonic Crystal (Au on Indium Tin Oxide) | Visible | Reflective. Strong resonant absorption. | Ultrafast phase modulation and polarization conversion. | Hundreds of fs | Ultrafast hot-electron transfer from plasmonic metal into ITO. |
| **Yang (2017)** [43] | Cadmium Oxide (CdO) Perfect Absorber | Near-IR (2.08 µm) | Reflective. p-polarized reflectance of 1.0% (near-perfect absorption), rising to 86.3% under pumping. | Reflective Polarizer: Switches ON/OFF with an extinction ratio of 91. | Sub-ps (switched on and off within 800 fs) | Transient increase of the effective electron mass of CdO under sub-bandgap pumping. |
| **Nicholls (2017)** [48] | Plasmonic Au Nanorod Array (Hyperbolic Metamaterial) | Visible | Transmissive (45° incidence). Strongly reduced output amplitude (hyperbolic regime). | Macroscopic Shift: Up to 60° rotation of the polarization ellipse. | Picosecond (≈3.4 ps) | Free-electron nonlinearities under control illumination (pump–probe) and in a self-phase modulation regime. |
| **This Work** | Au Orthogonal Nanorod Dimer Metasurface | Vis-NIR (700–850 nm) | Transmissive, normal incidence. High throughput (≈40% absolute transmission for a 45° probe, changing by ≤5 percentage points with one pump). | Transmissive Shift: ≈15° optical rotation and ≈10° ellipticity with one pump (45° probe); ≈22° and ≈15° with a second, delayed pump. The probe polarization selects intensity, ellipticity or rotation modulation (e.g. ≈25° rotation for a probe at 27°). | Rise within ≈100 fs; recovery over a few ps. A second pump can cancel the intensity change within ≈60 fs while the polarization stays modulated. | Transient anisotropy via selective, sequential non-equilibrium hot-electron heating of two independent resonators. |

**Table 1.** Comparative analysis of state-of-the-art ultrafast all-optical polarization control platforms. The performance metrics of the proposed Au orthogonal nanorod dimer metasurface are benchmarked against recent literature, highlighting the fundamental trade-offs between modulation depth (optical rotation and ellipticity), temporal recovery speed, and absolute transmissive efficiency across various material architectures and driving mechanisms. AZO: aluminium-doped zinc oxide; ENZ: epsilon-near-zero; IR: infrared; ITO: indium tin oxide; SHG: second-harmonic generation; TE: transverse electric; TM: transverse magnetic; Vis-NIR: visible to near-infrared.

## Discussion and Conclusions

In summary, our simulations show a plasmonic metasurface platform capable of ultrafast polarization control through selective, all-optical asymmetric perturbation, by utilizing an orthogonal nanorod dimer and the transient thermo-modulatory optical nonlinearities of Au. Unlike tunable metamaterials in which optical pumping modulates all modes uniformly, or transiently breaks the symmetry of an initially isotropic meta-atom[45], this design enables independent, pathway-specific manipulation of two spatially separate modes. The key element is the separation of the two orthogonal resonances into independent resonators, either of which can be heated selectively by choosing the pump polarization: the pumped resonator then acts as an all-optically tunable polarization channel, while the other remains an essentially unperturbed phase reference. The probe polarization selects which property of the transmitted light the pump modulates (**Figure 5**), and two delayed pumps address the two channels in sequence (**Figure 6**). Supported by full-wave FDTD simulations and a time-resolved 3TM, the results of this work show that polarization-selective pumping on orthogonally oriented dimer meta-atoms drives highly localized, non-equilibrium electron thermalization. This selective heating acts as a transient anisotropic driver, dynamically altering the relative amplitude and phase retardation between orthogonal localized surface plasmon modes. This active mode-mixing yields polarization modulation that rises within ≈100 fs, with ellipticity changes of ≈10° and optical rotations of ≈15° for a 45° probe; a second, delayed pump raises these to ≈15° and ≈22° and can cancel the intensity change of the first within ≈60 fs.

Crucially, this pronounced polarization control occurs while the transmittance changes by at most ≈0.05 (**Figure 3i**). By maintaining a nearly constant transmission of ≈40% alongside these transient polarization shifts, the proposed design offers a favourable performance profile for transmissive linear metasurfaces. While a recent all-dielectric metasurface[32] has reported larger absolute modulations, a phase retardation of up to 90° that turns linear into circular polarization, it relies on a reflective geometry, which provides no transmitted signal and complicates cascading in dense photonic networks. Conversely, existing transmissive platforms have either produced small transient anisotropies[19,45] or achieved macroscopic rotations only under demanding conditions: at oblique incidence and mJ $cm^{-2}$ control fluences with strong intensity modulation[46], in hyperbolic metamaterials at oblique incidence with a strongly reduced output amplitude[48], or through inefficient frequency conversion processes that reduce the transmitted signal[47].

Experimentally, the metasurface requires gold nanorods that can be fabricated by electron-beam lithography, and pump and probe pulses from standard amplified femtosecond lasers. As shown in the Supporting Information (Sections S1, S4 and S5), a 120-nm nanoporous film on a $CaF_2$ heat sink largely reproduces the computed response, the Kerr response of the backing wafer adds less than 1° of retardation, probe pulses of ≤70 fs resolve the sub-picosecond rise, and a rescaled design on a bulk substrate yields $\Delta\eta \approx 11°$ and $\Delta\alpha \approx 16°$ at ≈240 μJ/cm².

By delivering macroscopic polarization manipulation directly within the transmitted beam, this active mode-mixing approach avoids both the structural constraints of reflective devices and the strong optical attenuation of earlier transmissive platforms. While future experimental realizations will necessitate considerations for morphological damping inherent to nanofabrication and thermal management at high repetition rates, the proposed design provides a practical theoretical blueprint for ultrafast all-optical control. It eases the trade-offs that limit existing architectures, avoiding the strong transmissive attenuation of ENZ materials, the slow recovery of structural phase-change devices, and the low signal throughput of nonlinear frequency-conversion techniques. Beyond a single switch, the design offers two further handles for applications. Because the pump polarization selects which nanorod is heated, and each nanorod modulates the probe mainly near its own resonance (the two bands overlap only at ≈760–790 nm; **Figure 6d**; Supporting Information, Section S8), the polarization of the blue and red parts of a broadband probe can be modulated largely independently and at different times: wavelength-multiplexed polarization control set by the polarization and timing of the pumps. The probe polarization, in turn, sets whether the same metasurface acts as an intensity modulator, an ellipticity modulator or a polarization rotator (**Figure 5**). Combined, the second pump can end the intensity modulation of the first within ≈60 fs, so that the duration of an all-optical intensity gate is set by the pump delay rather than by the picosecond relaxation of the hot electrons, while the polarization stays modulated (**Figure 6e**). As a thin, in-line element at normal incidence that switches on within ≈100 fs and stays active for a few picoseconds, it could also serve as an all-optical polarization gate between polarizers, analogous to an optical Kerr gate, for time-resolved polarimetry and spectroscopy. Ultimately, this approach may provide a stepping stone towards all-optical polarization switches and modulators for nanophotonic circuits.

**Author information**

Corresponding Author N. Matthaiakakis[1,2*]

E-mail: n.matthaiakakis@inn.demokritos.gr

**Disclosures**

The authors declare no conflicts of interest.

**Acknowledgements**

The research project is implemented in the framework of the Hellenic Foundation for Research and Innovation (H.F.R.I.) call "4th Call for H.F.R.I.'s Research Projects to Support Postdoctoral Researchers" (H.F.R.I. Project Number: 28197). The authors gratefully acknowledge the Flexcompute EM Premium Education Program for providing complimentary access to the Tidy3D simulation tools used in this research. We would also like to acknowledge the valuable discussions with Dr. Andrea Schirato that were very useful in correctly replicating the 3TM.

# Supporting Information :

# Ultrafast All-Optical Polarization Control in a Plasmonic Metasurface by Selective Hot-Electron Heating of Orthogonal Gold Nanorods

*Nikolaos Matthaiakakis,[1,2]* Léa Testé[2], George Kakarantzas[2]*

[1] *Institute of Nanoscience and Nanotechnology, National Center for Scientific Research "Demokritos", 15341 Ag. Paraskevi, Greece*

[2] *Theoretical and Physical Chemistry Institute, National Hellenic Research Foundation, NHRF, 48 Vassileos Constantinou Ave., 11635 Athens, Greece*

**Corresponding author: n.matthaiakakis@inn.demokritos.gr*

This document provides supplementary information to "Ultrafast All-Optical Polarization Control in a Plasmonic Metasurface by Selective Hot-Electron Heating of Orthogonal Gold Nanorods".

Content

## S1. Computational Methodology

The linear optical properties and the ultrafast transient response of the Au nanorod dimer are investigated using full-wave 3D Finite-Difference Time-Domain (FDTD) simulations (Tidy3D).

### Static Geometrical Setup

In the simulations, the metasurface is defined by a unit cell with periodic boundary conditions along the x- and y-directions, featuring a periodicity of 500 nm. The unit cell comprises two discrete, orthogonally oriented Au nanoantennas resting on a semi-infinite dielectric substrate with a refractive index of n = 1.2. Both nanoantennas share a uniform thickness of 100 nm and are strictly co-planar. The y-oriented antenna (60 × 170 nm) is displaced by +100 nm from the unit cell centre along the y-axis, while the x-oriented antenna (210 × 60 nm) is displaced by -100 nm along the y-axis. The computational domain is terminated with Perfectly Matched Layers (PMLs) along the z-direction, placed sufficiently far from the metasurface to prevent artificial reflections. A high-resolution sub-grid mesh is enforced in the immediate vicinity of the gold structures to accurately resolve the highly confined near-fields of the localized surface plasmon (LSP) modes. In the layer of the nanorods, the cells measure 10 nm in x and y and 5 nm in z; on this mesh, the transmission dips of $LSP_x$ and $LSP_y$ lie at 807 and 734 nm.

### Multiphysics Simulation Pipeline

Because the ultrafast optical response of gold in the visible-to-near-infrared regime is governed by extreme non-equilibrium electron dynamics, the transient response is calculated using a unidirectional coupled multiphysics approach. The pipeline is executed in the following sequence:

1. **Static Pump Simulation:** First, a static FDTD simulation is performed to determine the absorbed power density ($P_{abs}$). The structure is illuminated at normal incidence by an x-polarized pump pulse (λ= 800 nm, azimuth angle $\varphi_{pu} = 0°$). The absorption profile within the excited Au nanorod is extracted to act as the driving source for the thermodynamic model.

2. **Three-Temperature Model (3TM):** The extracted $P_{abs}$ is then used to solve the coupled differential equations of the 3TM. This yields the temporal evolution of the non-thermal energy density ($N$), the thermal electron temperature ($\Theta_e$), and the lattice temperature ($\Theta_l$) following the femtosecond excitation.

3. **Transient Permittivity Calculation:** Using the time-dependent temperatures derived from the 3TM, the transient complex dielectric permittivity of gold, $\Delta\varepsilon(\omega, t)$, is analytically calculated. This calculation accounts for both the modification of the intraband Drude damping rate (due to increased electron-electron and electron-phonon scattering) and the smearing of the Fermi-Dirac distribution, which alters the X- and L-point interband transitions.

4. **Dynamic Probe Simulation:** Finally, the explicitly calculated transient permittivity ($\Delta\varepsilon$) is saved and imported back into the FDTD solver to update the material properties of the pumped gold nanorod. A broadband, linearly polarized probe pulse (azimuth angle $\varphi_{pr} = 45°$) is then simulated at various pump-probe delay times ($\Delta t$). The time-resolved transmission (Tr) is captured in the far field, from which the dynamic polarization parameters, ellipticity ($\eta$) and optical rotation ($\alpha$), are extracted.

This sequential methodology is justified by two primary assumptions in accordance with previously published works[1]:

- **Slowly Varying Envelope Approximation (SVEA):** The temporal duration of the pump pulse encompasses a sufficient number of optical cycles, making the SVEA valid for optical calculations.

- **Linear Absorption Regime:** We assume the pump fluence is sufficiently low to neglect nonlinear optical interactions within the pump pulse. While modelling femtosecond laser absorption in the purely linear regime is an approximation, recent literature indicates that this primarily results in a minor quantitative deviation in total absorbed energy, rather than altering the fundamental ultrafast dynamics (provided the fluence is moderate)[1].

**Substrate Modelling Approximations and Experimental Considerations**

a. Parasitic Nonlinear Substrate Effects (Optical Kerr Effect)

In our simulations, the substrate is modelled as an idealized linear medium (n = 1.2, representative of engineered nanoporous silica thin films). This low-index environment enhances near-field confinement and inherently suppresses parasitic Kerr nonlinearities due to the reduced material density. In physical experiments, this porous film would be deposited onto a transparent backing wafer (such as sapphire or $CaF_2$ for optimal thermal management). The Au nanorods can be fabricated by electron-beam lithography, and the femtosecond pump pulses at the two resonances and the broadband probe can be provided by standard amplified femtosecond lasers with optical parametric amplification or white-light generation. To validate the experimental viability of this configuration, full-wave dynamic 3TM-FDTD simulations were performed to evaluate the required layer thickness. As shown in Figure S1, an ultrathin porous layer of just 120 nm yields transient optical responses that largely follow those of the idealized semi-infinite $n = 1.2$ environment: the differential transmission (ΔTr) and the ellipticity change ($\Delta\eta$) differ by at most 0.0082 and 4.3° (on average by 0.002 and 0.25°). The backing wafer sharpens the static transmission band edge near 720 nm, where the first diffraction order of the 500-nm lattice starts to propagate in the $n = 1.45$ wafer (1.45 × 500 nm ≈ 725 nm), and the narrow optical-rotation feature near 770 nm, where the transmitted light is nearly circular and $\Delta\alpha$ reaches ≈34° instead of ≈21°. A film thickness of ≥ 120 nm is therefore sufficient to contain the localized plasmonic near fields of the Au nanorods: optically, the nanorods behave almost as on a bulk low-index substrate, and the transient polarization changes are preserved. An experimental target thickness of 120–250 nm thus offers a practical tolerance window: ≈120 nm provides near-complete optical isolation while minimizing the Kapitza resistance and thermal barrier to the backing sapphire or $CaF_2$ wafer.

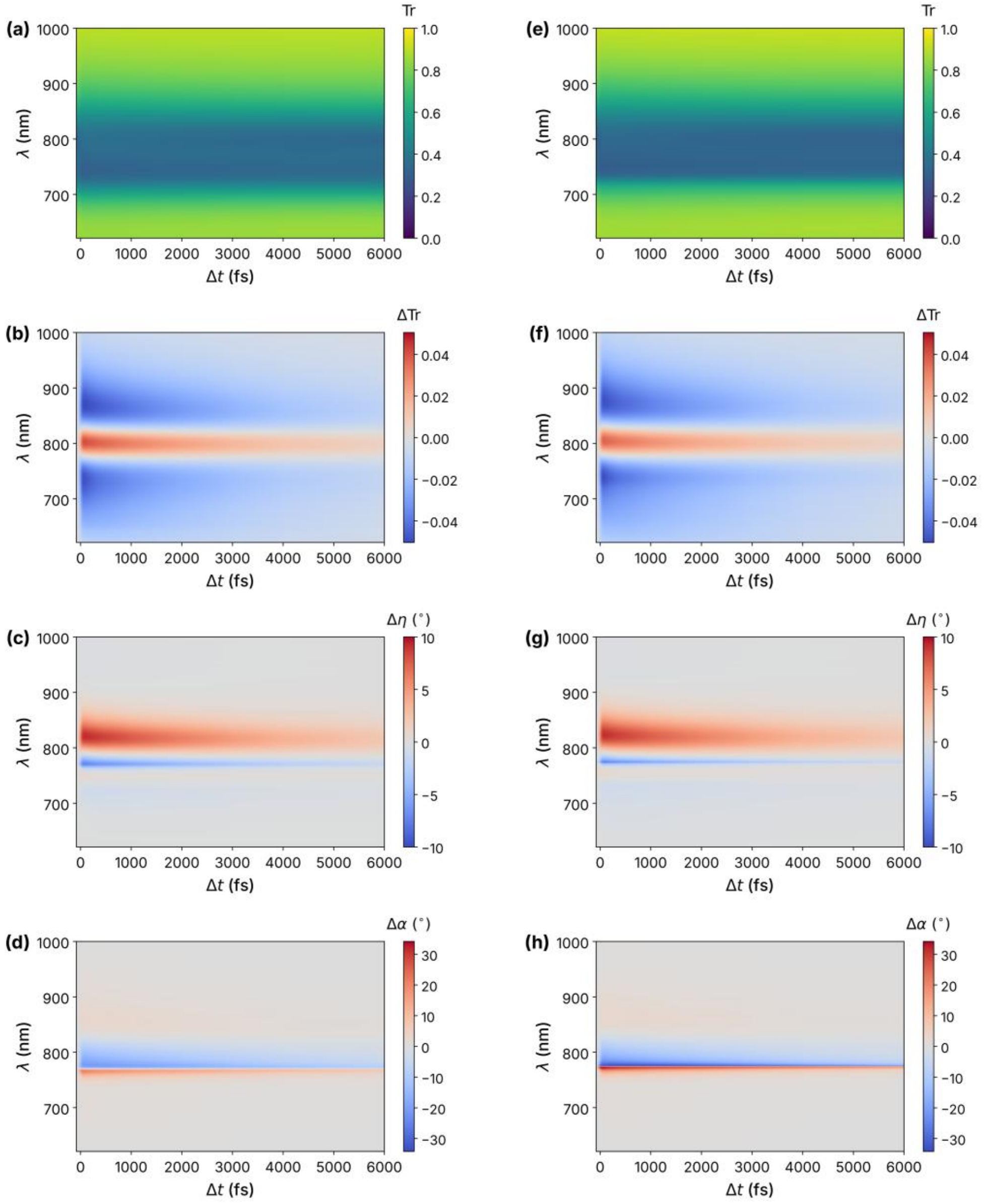

**Figure S1. Transient optical response comparison between bulk and ultrathin-film substrate configurations.** Simulated pump-induced transient modulations mapped as a function of probe wavelength ($\lambda$) and pump-probe delay ($\Delta t$). The left column **(a–d)** displays the dynamic response of the orthogonal Au nanorod dimers situated on an idealized, semi-infinite low-index substrate ($n = 1.2$). The right column **(e–h)** displays the identical metasurface situated on an ultrathin 120 nm porous film ($n = 1.2$) deposited over a standard higher-index transparent backing substrate ($n = 1.45$, representative of a $CaF_2$ heat sink). The rows depict the absolute transmission **(a, e)**, differential transmission **(b, f)**, differential ellipticity **(c, g)**, and differential optical rotation **(d, h)**. Each row uses one colour scale

for both columns. The ΔTr and $\Delta\eta$ maps are similar in the two configurations (differences of 0.002 and 0.25° on average, at most 0.0082 and 4.3°).

For standard bulk backing substrates (e.g., 500 μm $CaF_2$), the pump induces transient cross-phase modulation and birefringence via the instantaneous optical Kerr effect. Given the nonlinear refractive index of $CaF_2$ ($n_2 \approx 1.9 \times 10^{-16}$ cm²/W), [2] and our peak pump intensity ($I \approx 1.1 \times 10^{10}$ W/cm² for a 70-fs Gaussian pulse at 840 μJ/cm²), the pump-induced birefringence for a probe polarized at 45° to the pump is $\Delta n = (4/3)n_2 I \approx 2.9 \times 10^{-6}$. Across 500 μm, this generates a parasitic phase retardation below 1° (approximately 0.6°). Crucially, because this instantaneous response is strictly confined to the ≈70 fs pump-probe overlap window, its impact on the broader metasurface modulation dynamics is minimal.

Furthermore, the localized plasmonic near-fields are dominated by the immediate nanoporous dielectric environment rather than the backing substrate, further isolating the measured response from bulk substrate nonlinearities. Any residual artifacts are easily eliminated experimentally using standard baseline subtraction, non-collinear pump-probe geometries, or ultra-thin membrane substrates (e.g., 50 nm SiN).

b. Decoupling of Ultrafast Thermalization from Substrate Heat Dissipation

The 3TM computes the internal thermodynamic evolution of the gold nanorods while intentionally neglecting thermal diffusion into the surrounding substrate. This approximation is justified by the stark separation of timescales inherent to nanoscale heat transport. The internal electron-phonon thermalization responsible for the transient optical anisotropy and macroscopic dielectric modulation occurs within a few picoseconds. Conversely, heat transfer from the gold lattice to the dielectric substrate is bottlenecked by the interfacial Kapitza resistance, unfolding over tens to hundreds of picoseconds. Because the peak polarization shifts reported here operate within the sub-picosecond regime (peaking within ≈100 fs), the substrate effectively acts as a thermal insulator during the active switching window. The bulk thermal conductivity of the substrate only becomes physically relevant on nanosecond timescales during the long-term cooling phase; here, higher-conductivity substrates like $CaF_2$ can be strategically utilized to accelerate steady-state heat dissipation, thereby raising the boundary for the maximum allowable laser repetition rate.

**Pump fluence selection**

The incident pump fluence (≈840 μJ/cm²) balances the magnitude of the polarization modulation against the thermal load on the metasurface. The value used is well below that used in other theoretical works[3] and in the range of experimentally verified works [1]. Furthermore, because our architecture is designed to maintain a high absolute signal throughput (≈40%), only ≈22% of the pump is absorbed (Section S8), so that a somewhat higher incident fluence is required than in metasurfaces designed for strong absorption (e.g. [1]). The absorbed energy raises the lattice temperature by ≈130 K within 6 ps (Figure 3d of the main text) and by ≈150 K once electrons and lattice have equilibrated, far below the melting point of gold.

## S2. Three-temperature model

To quantitatively describe the ultrafast thermodynamic response of the gold nanoantennas following femtosecond laser excitation, a 3TM is employed[3–5]. When an intense, ultrashort pump pulse interacts with a metal, the absorbed photon energy initially creates a highly non-equilibrium, non-thermal distribution of electrons. The 3TM tracks the sub-picosecond energy relaxation through three coupled thermodynamic reservoirs: the non-thermal electron energy density (*N*), the thermalized electron gas temperature ($\Theta_e$), and the lattice temperature ($\Theta_l$). The temporal dynamics of these interacting energy subsystems are governed by the set of coupled differential equations:

$$\frac{dN(t)}{dt} = -\alpha_s N(t) - bN(t) + P_{abs}(t), \qquad (s1)$$

$$C_e \frac{d\Theta_e(t)}{dt} = -G\big(\Theta_e(t) - \Theta_l(t)\big) + \alpha_s N(t), \qquad (s2)$$

$$C_l \frac{d\Theta_l(t)}{dt} = G\big(\Theta_e(t) - \Theta_l(t)\big) + bN(t), \qquad (s3)$$

where $P_{abs}$ is the absorbed power density from the pump pulse, $\alpha_s$ and $b$ are the scattering rates defining non-thermal energy transfer to the electron gas and lattice respectively, and $C_e$ and $C_l$ are the electron gas and lattice heat capacities. The lattice heat capacity ($C_l = 2.49 \times 10^6$ J/(m$^3$ K)) remains approximately constant over the temperature range of interest[1]. Conversely, the electron heat capacity ($C_e$) exhibits a strong linear dependence on the electronic temperature, defined as $C_e(\Theta_e) = \gamma_e \Theta_e$ [4], where $\gamma_e$ is the electron heat capacity coefficient. $G$ is the electron-phonon coupling constant[1]:

$$G(\Theta_e) = \frac{\pi \hbar k_B \Lambda \langle \omega_{ph}^2 \rangle}{D(E_F)} \int_{-\infty}^{\infty} D^2(E) \left(-\frac{\partial f}{\partial E}\right) dE \quad (s4)$$

where $\Lambda$ is the electron-phonon coupling factor specific for Au, $\langle \omega_{ph}^2 \rangle$ is the average value of the square of the phonon frequency with $\Lambda \langle \omega_{ph}^2 \rangle = 23 \pm 4 \quad \text{meV}^2$, $E_F$ is the Fermi energy of gold, $D(E)$ is the density of states[6], and $f$ is the Fermi-Dirac distribution.

The system is driven by the source term $P_{\text{abs}}(\mathbf{t})$, which defines the instantaneous absorbed power density injecting energy into the non-thermal electron population. This term is calculated as:

$$P_{\text{abs}}(t) = \frac{FS}{V} A g(t) \quad (s5)$$

In this expression, $F$ denotes the incident pump fluence, $S$ represents the area of the metasurface unit cell, and $V$ is the volume of the pumped nanorod, $A$ the absorption by the nanostructure. Finally, $g(t)$ defines the temporal intensity envelope of the ultrashort pump pulse, modelled as a normalized Gaussian function:

$$g(t) = \sqrt{\frac{4 \ln 2}{\pi \Delta t_{pulse}{}^2}} \exp\left[-\frac{4 \ln 2 (t - t_0)^2}{\Delta t_{pulse}{}^2}\right] \quad (s6)$$

Here, $\Delta t_{pulse}$ represents the pulse duration defined as the full-width at half-maximum (FWHM). An effective pulse duration of $\Delta t_{pulse} = 70$ fs was utilized in the simulations. The coupled differential equations (s1)–(s3) governing the temporal dynamics were solved numerically via a standard fourth-order Runge-Kutta (RK4) method.

The parameter $\alpha_s$ represents the average heating rate at which the highly energetic, non-thermal electrons scatter and thermalize into a hot Fermi-Dirac distribution. Based on Fermi liquid theory, this scattering rate scales quadratically with the excitation energy. Thus, we define the rate as $\alpha_s = \langle 1/\tau_{e-e} \rangle$, given by[3]:

$$\alpha_s = \frac{(\hbar \omega_{\text{pump}})^2}{2 \tau_0 E_F^2} \quad (s7)$$

where $\hbar\omega_{\text{pump}}$ is the pump photon energy, and $\tau_0 = 7\,fs$ an empirical scattering time constant characteristic of the metal.

Concurrently, the parameter $b$ governs the direct energy transfer rate from the non-thermal electron population to the lattice (phonons). This out-of-equilibrium decay rate is defined as $b = \langle 1/\tau_{e-ph} \rangle$, expressed as:

$$b = \frac{k_B \Theta_D}{\tau_f \hbar\omega_{\text{pump}}} \quad (s8)$$

where $k_B$ is the Boltzmann constant, $\Theta_D = 165\,K$ is the Debye temperature of gold, and $\tau_f = 13.824\,fs$ is the quasi-particle free flight time[1]. Together, these terms accurately capture the transient decay of the non-thermalized distribution before the system fully converges into the coupled thermal dynamics of $\Theta_e$ and $\Theta_l$.

**S3. Permittivity changes**

Once the temporal dynamics of $N$, $\Theta_e$, and $\Theta_l$ are determined via the 3TM, the corresponding photogenerated modulation of the complex permittivity is computed by evaluating both intraband and interband contributions [1].

**Intraband Contribution:** In this modelling framework, the intraband contribution is expressed using a Drude term. This term is modified by changes in both electron and lattice temperatures through the Drude damping Γ and plasma frequency $\omega_p$. For energy-loss processes involving electron scattering, the total scattering rate can be expressed as the sum of electron-electron and electron-phonon contributions [1]:

$$\Gamma = \Gamma_{e-e} + \Gamma_{e-ph} \quad (s9)$$

The electron-electron scattering rate, derived from Landau's Fermi liquid theory for electron gas at temperature $\Theta_e$, is given by:

$$\Gamma_{e-e}(\hbar\omega, \Theta_e) = \frac{(k_B\Theta_e)^2}{\hbar^2\omega_p}\left[1 + \left(\frac{\hbar\omega}{2\pi k_B \Theta_e}\right)^2\right] \quad (s10)$$

where $\hbar\omega$ represents the photon energy. The electron-phonon scattering rate, assuming parabolic conduction bands and negligible phonon energies compared to photon energies at optical frequencies, is expressed as:

$$\Gamma_{e-ph}(\hbar\omega, \Theta_l, \Theta_e) = \gamma_{e-ph}\frac{\Theta_l}{\hbar\omega}\int_0^\infty \sqrt{E}\sqrt{E+\hbar\omega}f(E,\Theta_e)[1-f(E+\hbar\omega,\Theta_e)]dE \quad (s11)$$

Here, $E$ is electron energy, $f(E,\Theta_e)$ is the Fermi-Dirac distribution:

$$f(E,\Theta_e) = \frac{1}{1+\exp\left(\frac{E-E_F}{k_B\Theta_e}\right)} \quad (s12)$$

where $E$ is the absolute electron energy, and $\Theta_0$ is the initial electron temperature. $\gamma_{e-ph} \approx 7 \times 10^{10}$ rad $s^{-1}eV^{-1}K^{-1}$ is a fitting constant. For the electron temperature range in this study, a simplified expression can be used:

$$\Gamma_{e-ph}(\hbar\omega, \Theta_l) \simeq \Gamma^0_{e-ph}(\hbar\omega) + \beta_{e-ph}(\hbar\omega)^2\Delta\Theta_l \quad (s13)$$

where the linearization constant is $\beta_{e-ph} = 1.95 \times 10^{48}$ $s^{-1}J^{-2}K^{-1}$. The equilibrium electron-phonon scattering rate is computed from the full integral expression by setting the electron and lattice temperatures equal to the ambient temperature $\Theta_l = \Theta_e = \Theta_0 = 300$ K. The plasma frequency contribution to permittivity modulation arises from thermal expansion of the metal structure, which affects free electron density in the conduction band. The particle volume changes with temperature with a volume thermal expansion coefficient $\alpha_V(\Theta_e) = \frac{1}{B}\left[\lambda_G C_l + \frac{2}{3}C_e(\Theta_e)\right]$ with $B = 180$ GPa being the bulk modulus and $\lambda_G = 2.95$ the Grüneisen parameter. The temperature-dependent plasma frequency becomes:

$$\omega_p(\Theta_e, \Theta_l) = \sqrt{\frac{e^2}{\varepsilon_0 m^*}n(\Theta_e, \Theta_l)} \quad (s14)$$

where $e$ is electron charge, $\varepsilon_0$ is vacuum permittivity, m* is effective electron mass, n is conduction electron density.

**Interband Contribution:** The interband contribution to the dielectric function arises from direct electronic transitions between the occupied d-band states and unoccupied sp-band states above the Fermi level. In gold, the dominant interband transitions occur at the X and L high-symmetry points of the Brillouin zone [6,7].

Following the formulation of Marini et al.[8], the imaginary part of the interband permittivity is expressed as:

$$\varepsilon_{inter}''(\omega, \Theta_e) = \frac{\pi e^2}{3\varepsilon_0 m_0^2 \omega^2}[|\tilde{p}_{cv}^X|^2 J_{cv}^X(\omega, \Theta_e) + |\tilde{p}_{cv}^L|^2 J_{cv}^L(\omega, \Theta_e)] \quad (s15)$$

where $e$ is the electron charge, $\varepsilon_0$ is the vacuum permittivity, $m_0$ is the free electron mass, $\omega$ is the angular frequency, $|\tilde{p}_{cv}|^2$ are the squared dipole matrix elements, and $J_{cv}(\omega, \Theta_e)$ is the temperature-dependent joint density of states for each transition. The squared dipole matrix elements characterize the optical transition strength at each symmetry point. Following Ref. [8], the L-point matrix element is: $|\tilde{p}_{cv}^L|^2 = 2.00 \times 10^{-48}$ J · kg. The ratio of oscillator strengths between X and L transitions was determined by Guerrisi et al. [6] from analysis of the interband absorption edge splitting in gold $\frac{|\tilde{p}_{cv}^X|^2}{|\tilde{p}_{cv}^L|^2} = 0.321 \times \frac{g_L}{g_X}$ where $g_L = 8$ and $g_X = 6$. The larger value for the L transition reflects its dominant contribution to interband absorption in gold near 2.5 eV. This can be expressed directly in terms of the change in joint density of states (JDOS) which accounts for the number of allowed transitions at a given photon energy, weighted by the Fermi-Dirac occupation factors. Following the parabolic band approximation [4,7,9]:

$$\Delta\varepsilon_{inter}''(\omega) = \frac{\pi e^2}{3\varepsilon_0 m_0^2 \omega^2}[|\tilde{p}_{cv}^X|^2 \Delta J_X(\omega) + |\tilde{p}_{cv}^L|^2 \Delta J_L(\omega)] \quad (s16)$$

where[4]:

$$\Delta J(\omega) = \int_{E_{min}}^{E_{max}} [f(E, \Theta_e) - f(E, \Theta_0)]\, D(E)\, dE \quad (s17)$$

The occupation of electronic states is governed by the Fermi-Dirac distribution. For the evaluation of the interband transitions, it is computationally convenient to perform a change of variables such that the electron energy $E$ is defined relative to the Fermi level ($E = 0$ at $E_F$). The integration over final states extends from $E_{min} = -20k_B\Theta_0 \approx$-0.52 eV to an upper limit that depends on photon energy and band structure. For the X transition:

$$E_{max,X} = \hbar\omega_{X_6^-} + \frac{(\hbar\omega_{X_7} + \hbar\omega_{X_6^-} - \hbar\omega)m_{u,X}^{II}}{m_{l,X}^{II} - m_{u,X}^{II}} \quad (s18)$$

and for the L transition:

$$E_{max,L} = \hbar\omega_{L_4^-} + \frac{(\hbar\omega_{L_4^-} + \hbar\omega_{L_4} - \hbar\omega)m_{u,L}^{II}}{m_{l,L}^{II} - m_{u,L}^{II}} \quad (s19)$$

where $\hbar\omega_{X_7} + \hbar\omega_{X_6^-}$ and $\hbar\omega_{L_4^-} + \hbar\omega_{L_4}$ are the interband gap energies at the critical points.

The band structure near the X and L points is described using a parabolic approximation with effective masses determined from relativistic band calculations by Christensen and Seraphin[7,9]. The X-point transition involves the $X_6^-$ and $X_7$ bands. The relevant effective masses are: $m_{l,X} = 0.770m_0$, $m_{l,X}^{II} = 0.814m_0$ and $m_{u,X} = 0.220m_0$, $m_{u,X}^{II} = 0.180m_0$. The critical point energies defining the X transition are the gap $\hbar\omega_{X_7} + \hbar\omega_{X_6^-} = 3.125$ eV, and the upper final state band limit $\hbar\omega_{X_6^-} = 1.475$ eV.

The L-point transition involves the $L_4$ and $L_4^-$ bands. The effective masses are: $m_{l,L} = 0.720m_0$, $m_{l,L}^{II} = 0.774m_0$ and $m_{u,L} = 0.190m_0$, $m_{u,L}^{II} = 0.260m_0$. The critical point energies for the L transition are the gap $\hbar\omega_{L_4^-} + \hbar\omega_{L_4} = 2.292$ eV, and the upper final state band limit $\hbar\omega_{L_4^-} = 0.717$ eV.

Following the approach of Ref. [9], the non-thermal electron population (energy density *N*) generated by ultrafast laser excitation also modifies the interband response. The change in JDOS due to non-thermal carriers is[4]:

$$\Delta J_N(\omega) = \frac{N}{\alpha_{JN}} \int_{E_{min}}^{E_{max}} \left[ f\left(E - \hbar\omega_{pump}, \Theta_0\right)\left(1 - f(E, \Theta_0)\right) - f(E, \Theta_0)\left(1 - f\left(E + \hbar\omega_{pump}, \Theta_0\right)\right)\right] D(E)\, dE \quad (s20)$$

where $\alpha_{JN}$ is a scaling parameter relating the non-thermal population to the induced carrier density. The real and imaginary parts of the dielectric function are not independent but are connected through the Kramers-Kronig relations, which arise from causality requirements[1]. Thus, having computed the change in the imaginary part of the interband permittivity, the corresponding change in the real part is obtained via the Kramers-Kronig transformation.

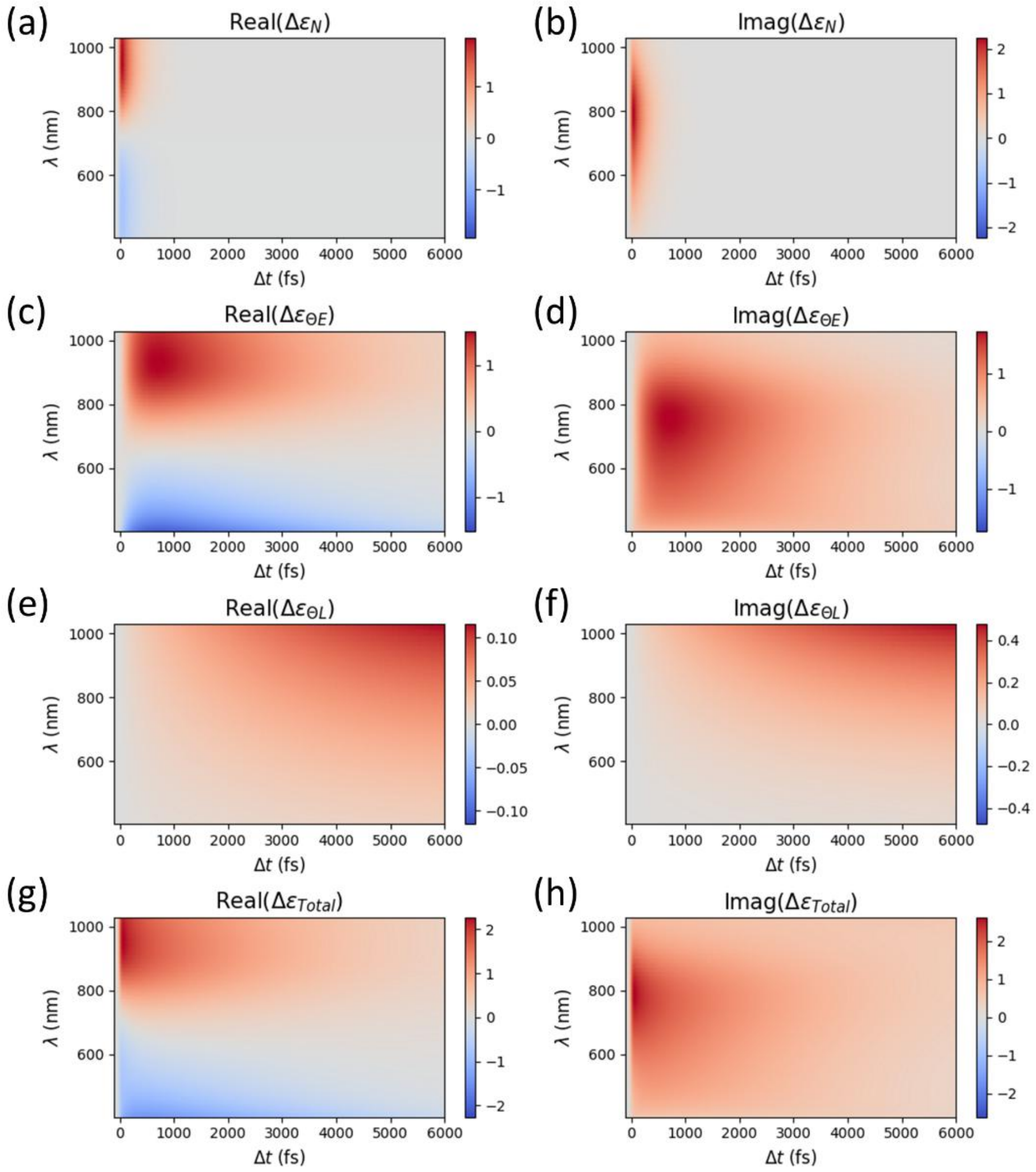

**Figure S2. Decomposed transient dielectric permittivity dynamics of gold.** Simulated maps detailing the distinct physical contributions to the pump-induced changes in the real (left column) and imaginary (right column) parts of the complex dielectric function ($\Delta\varepsilon$) as a function of probe wavelength ($\lambda$) and pump-probe delay ($\Delta t$). **(a, b)** Contribution from the non-thermal electron population (energy density $N$), characterized by an instantaneous, highly transient response that decays within a few hundred femtoseconds. **(c, d)** Contribution from the thermalized electron gas temperature ($\Theta_e$), which dominates after the first few hundred femtoseconds and drives the pronounced sub-picosecond permittivity shift primarily via Fermi smearing of the interband transitions. **(e, f)** Contribution from the lattice temperature ($\Theta_l$), exhibiting a significantly smaller magnitude (note the reduced scale of the colour bar) and a delayed onset that governs the long-tail picosecond relaxation dynamics as the hot electron gas transfers energy to the phonons. **(g, h)** The total transient permittivity change ($\Delta\varepsilon_{Total}$), representing the comprehensive sum of all three thermodynamic subsystems, which are ultimately imported into the dynamic FDTD simulations.

To elucidate the microscopic origin of the transient polarization control, the temporal and spectral evolution of the gold permittivity was decomposed into its constituent thermodynamic drivers (Figures S2 and S3). Upon optical excitation, the initial dielectric response is governed by the non-thermal electron population (energy density $N$), which induces an instantaneous but short-lived perturbation that decays within a few hundred femtoseconds (Figure S2a–b). As these highly energetic carriers rapidly interact, they form a thermalized hot-electron gas ($\Theta_e$) that dominates the optical nonlinearity after the first few hundred femtoseconds.

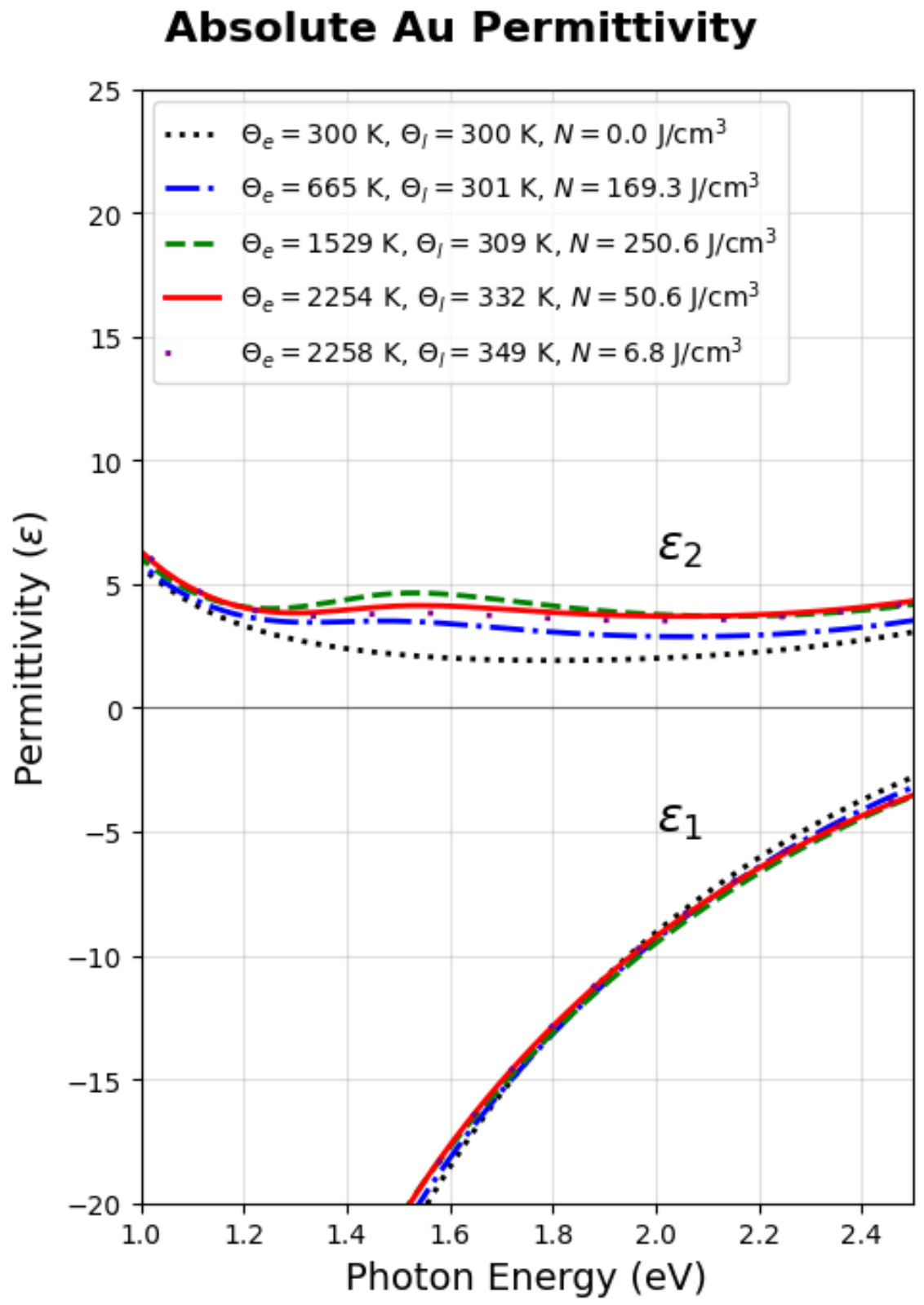

**Figure S3. Absolute complex permittivity of gold (Au) under varying transient thermodynamic conditions.** The graph displays the spectral dispersion of the real ($\varepsilon_1$) and imaginary ($\varepsilon_2$) components of the Au dielectric function as a function of photon energy. The black dotted line represents the material's static baseline at room-temperature equilibrium ($\Theta_e$ = 300 K, $\Theta_l$ = 300 K, $N$ = 0.0 J/cm³). The coloured curves illustrate the pronounced modulation of the complex permittivity at different stages of ultrafast optical excitation. These dynamic states are defined by the interplay of the thermal electron temperature ($\Theta_e$), lattice temperature ($\Theta_l$), and non-thermal electron energy density ($N$), capturing the progression from an initial surge in non-thermal electrons (at 0 and 100 fs) to extreme thermal electron heating (at 0.5 and 1 ps, close to the peak electron temperature of ≈2290 K at ≈0.7 ps).

As seen in Figure S2c and S2d, this $\Theta_e$ contribution drives a strong, sub-picosecond modulation of the dielectric function, primarily arising from Fermi-Dirac smearing around the interband transition. Finally, subsequent electron-phonon coupling transfers this energy to the lattice ($\Theta_l$), which contributes a comparatively weak, delayed background responsible for the long-tail picosecond relaxation (Figure S2e–f).

The absolute permittivity spectra (Figure S3) further reflect this temporal progression, illustrating pronounced shifts in both $\varepsilon_1$ and $\varepsilon_2$ as the electron temperature approaches its peak of ≈2290 K. Crucially, the spectral shape, magnitude, and temporal hierarchy of these extracted dynamics agree with established literature. The modulation of the complex permittivity under strong electronic excitation agrees with the high-temperature responses reported by Yurkevich et al. [10]. Furthermore, the temporal decomposition, specifically the rapid progression from $N$ to a dominant $\Theta_e$ regime, followed by a weak $\Theta_l$ background, is consistent with the work of Schirato et al.[1] for active plasmonic architectures. This comparison supports the material inputs of our dynamic FDTD framework.

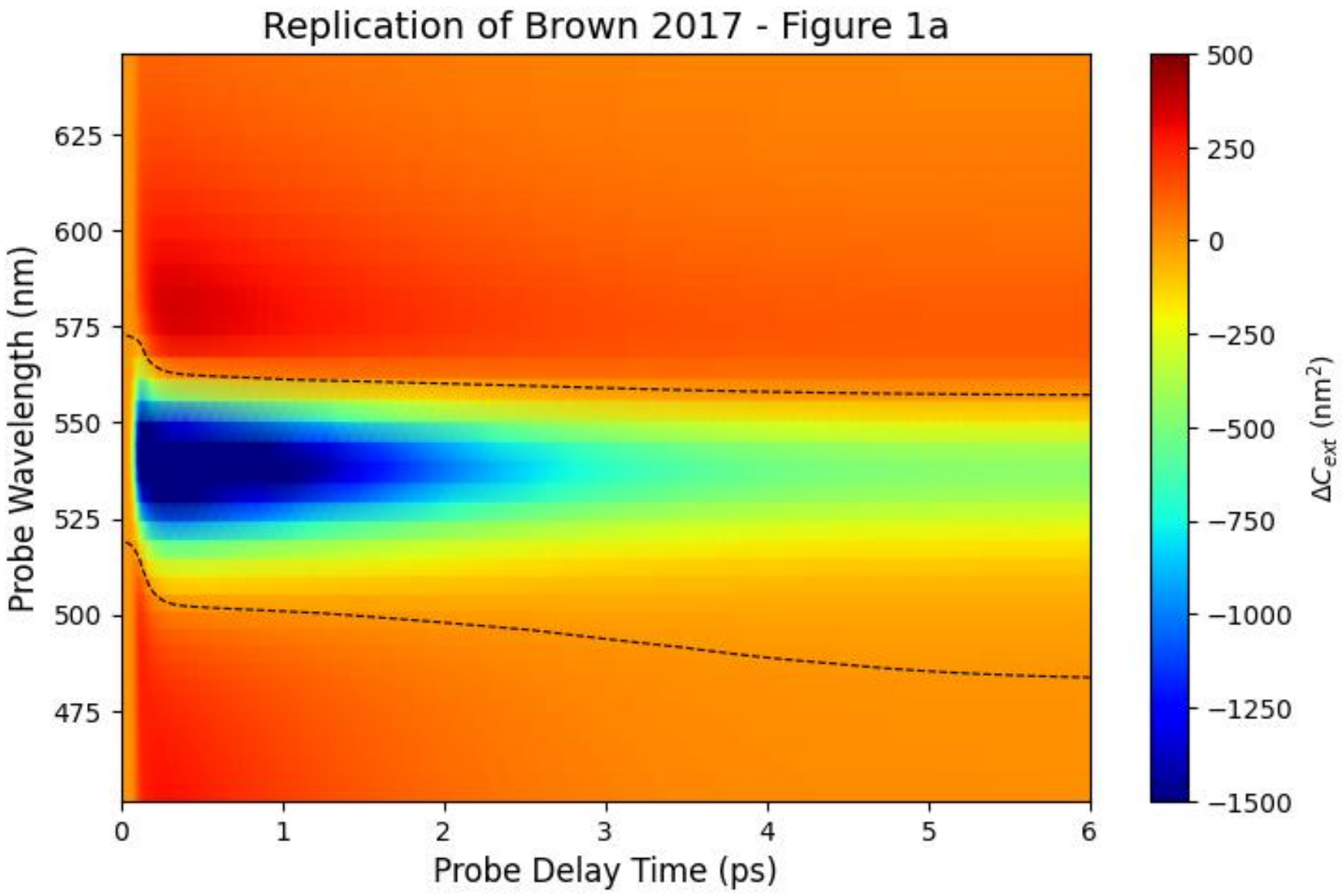

**Figure S4.** Simulation of the experiment of Brown et al. (2017; their Figure 1a) with the present model. This map displays the differential extinction cross section ($\Delta C_{ext}$) of colloidal gold nanoparticles as a function of probe delay

time and probe wavelength. As in the original study[11], this response was generated using a pump pulse with a fluence of 68 μJ/cm² at 380 nm. At a delay time of zero, the pump pulse excites the sample. As internal thermalization of the electrons occurs, the extinction near the 533 nm absorption peak decreases, resulting in a strong negative signal. Simultaneously, the extinction increases in the wings on either side of the absorption peak, corresponding to the positive signal regions. The dashed black contour lines represent the threshold of zero extinction change.

To further test our computational framework, we benchmarked our material models against the experimental and ab initio ultrafast carrier dynamics reported by Brown *et al.*[11]. Figure S4 presents our simulation of their transient extinction cross-section ($\Delta C_{ext}$) map for colloidal gold nanoparticles. Following optical excitation, the rapid thermalization of the electron gas induces a pronounced bleaching of the localized surface plasmon resonance (LSPR) near 533 nm, manifesting as the dominant negative $\Delta C_{ext}$ signal. This central bleach is immediately accompanied by positive transient absorption in the spectral wings, a classic signature of plasmon broadening driven by Fermi smearing of the interband transitions. The temporal evolution captured in our model, featuring a sub-picosecond onset during electron thermalization followed by a multi-picosecond relaxation tail governed by lattice heating, agrees with the experimental and ab initio results of the original study. Reproducing these spectral contours supports the use of our coupled 3TM–FDTD methodology to translate hot-carrier thermodynamics into plasmonic observables.

**S4. Influence of probe duration**

In theoretical pump-probe simulations, the calculated optical parameters typically represent an idealized, instantaneous material response. However, in physical transient polarimetry experiments, the measured signal is governed by the temporal cross-correlation between this intrinsic material response and the intensity envelope of the probe pulse. To assess the practical experimental observability of our results, we evaluated the effect of finite probe pulse durations on the transient mode-mixing dynamics.

The idealized, instantaneous transient response was convolved with Gaussian probe intensity envelopes of varying FWHM durations, ranging from 35 fs to 240 fs. As illustrated in Figure S5, the temporal smearing introduced by the finite probe width fundamentally alters the observed sub-picosecond dynamics. Specifically, longer probe pulses (e.g., 240 fs) artificially broaden the signal onset, delay the apparent peak time, and dampen the maximum achievable amplitude of the optical modulation (including $\Delta Tr$, $\Delta\eta$, and $\Delta\alpha$).

Conversely, the slower picosecond recovery tail, driven by electron-phonon thermalization, remains largely unaffected by the probe duration, as its timescale safely exceeds the pulse width. These findings underscore a critical experimental design parameter: to resolve the sub-picosecond rise of the polarization response of the proposed metasurface, deploying ultrashort probe pulses (≤ 70 fs) is important to prevent severe temporal attenuation of the signal.

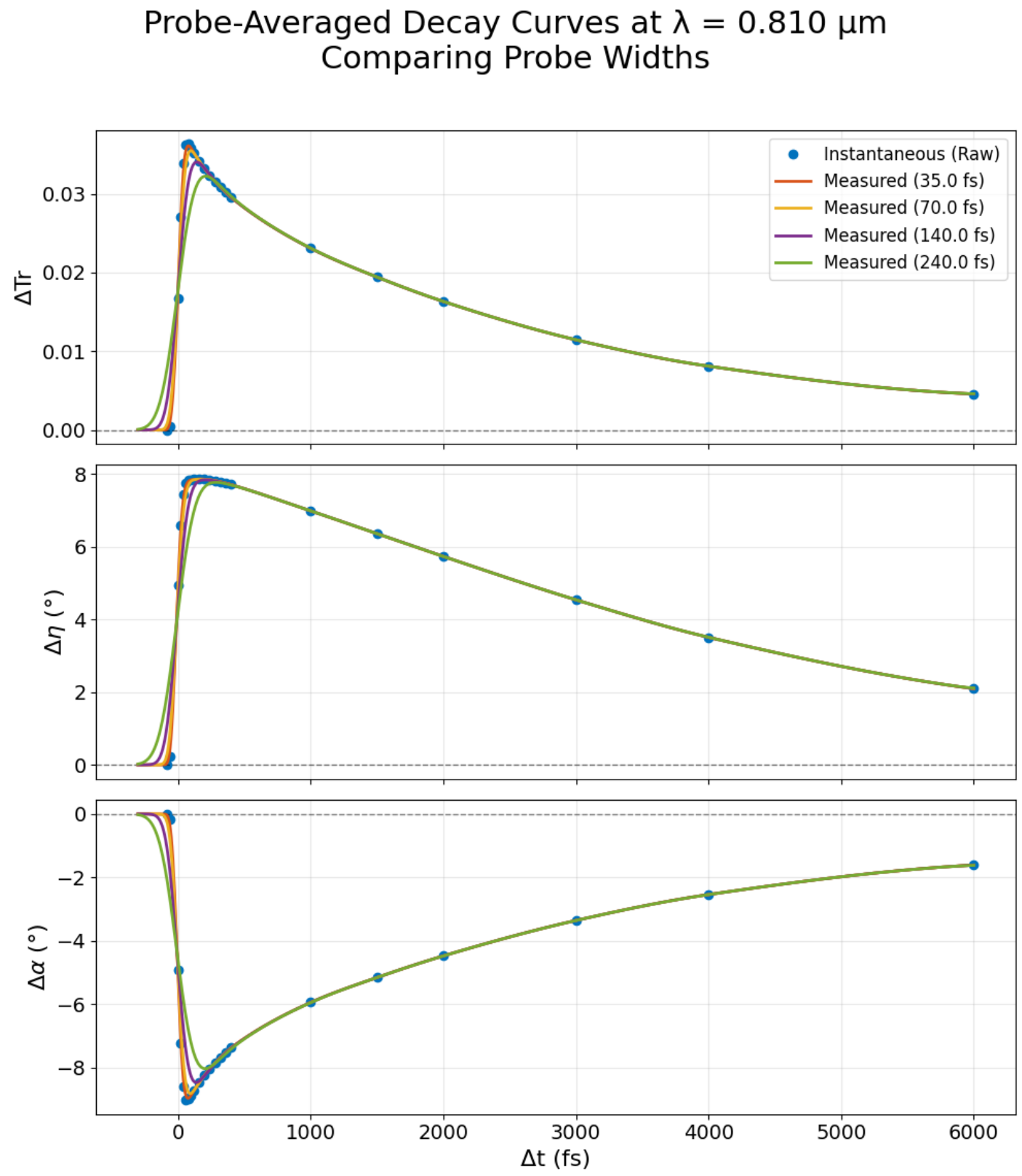

**Figure S5. Effect of temporal convolution with finite-duration probe pulses.** Simulated transient decay curves evaluated at a probe wavelength of λ = 810 nm for transient differential transmission (ΔTr), ellipticity ($\Delta\eta$), and optical rotation ($\Delta\alpha$). The instantaneous, raw material response (blue dots) is compared against signals convolved with Gaussian probe pulses of varying FWHM durations (35 fs to 240 fs). While ultrashort probes (35 fs, 70 fs) precisely track the highly nonlinear sub-picosecond onset and preserve the peak modulation depth, longer probes (140 fs, 240 fs) introduce temporal smearing, effectively dampening the peak amplitude and artificially broadening the ultrafast response time.

### S5. Metasurface Design Flexibility: High-Index Substrate and Scaled Geometry

As discussed in the main text, adapting the metasurface for high-repetition-rate experiments may necessitate fabricating the Au dimers directly on higher thermal conductivity bulk substrates, such as fused silica or $CaF_2$. To compensate for this higher refractive index environment ($n = 1.45$) and maintain optimal resonant excitation near the 800 nm pump wavelength, the geometry of the unit cell must be scaled.

This section illustrates the flexibility of the active mode-mixing framework by evaluating a modified nanorod dimer design on an $n = 1.45$ substrate, with the corresponding cross-sectional schematic shown in **Figure S6a**.

**Geometrical Scaling and Pumping Parameters:** To preserve the operating spectral window, the dimensions of the Au meta-atoms were systematically reduced:

- **Thickness:** The Au layer thickness was reduced from 100 nm to 50 nm.
- **y-oriented nanorod ($LSP_y$):** The length was reduced from 170 nm to 140 nm (width maintained at 60 nm).
- **x-oriented nanorod ($LSP_x$):** The length was reduced from 210 nm to 160 nm (width maintained at 60 nm).
- **Excitation:** The structure is pumped at 800 nm with a 70 fs pulse. Due to the reduced active volume and altered near-field confinement, a lower incident pump fluence of ≈240 μJ/cm² is utilized to drive the required hot-electron nonlinearity.

**Transient Optical Response:** As shown in **Figure S6**, the scaled design successfully replicates the core ultrafast polarization switching dynamics:

- **Thermodynamic Evolution (Figure S6b–d):** The reduced footprint and adjusted fluence still drive intense non-equilibrium electron heating. The thermal electron gas temperature ($\Theta_e$) peaks just above 2500 K, while the lattice ($\Theta_l$) gradually reaches ≈450 K.
- **Permittivity and Transmission (Figure S6e, f, i):** The strong change of the complex dielectric function yields a pronounced transient differential transmission (ΔTr).
- **Polarization Modulation (Figure S6g, h, j, k):** The asymmetric perturbation modulates the optical anisotropy of the unit cell. The structure delivers large, ultrafast polarization changes, reaching peak transient modulations of ≈11° in ellipticity (Δη) and ≈16° in optical rotation (Δα).

It should be emphasized that this scaled configuration serves as a preliminary proof-of-concept demonstration rather than an exhaustively optimized device. Nevertheless, these results indicate that the proposed orthogonal dimer architecture is adaptable: it can be engineered to accommodate various substrate environments and thermal management requirements while keeping the sub-picosecond rise of the response and large polarization changes.

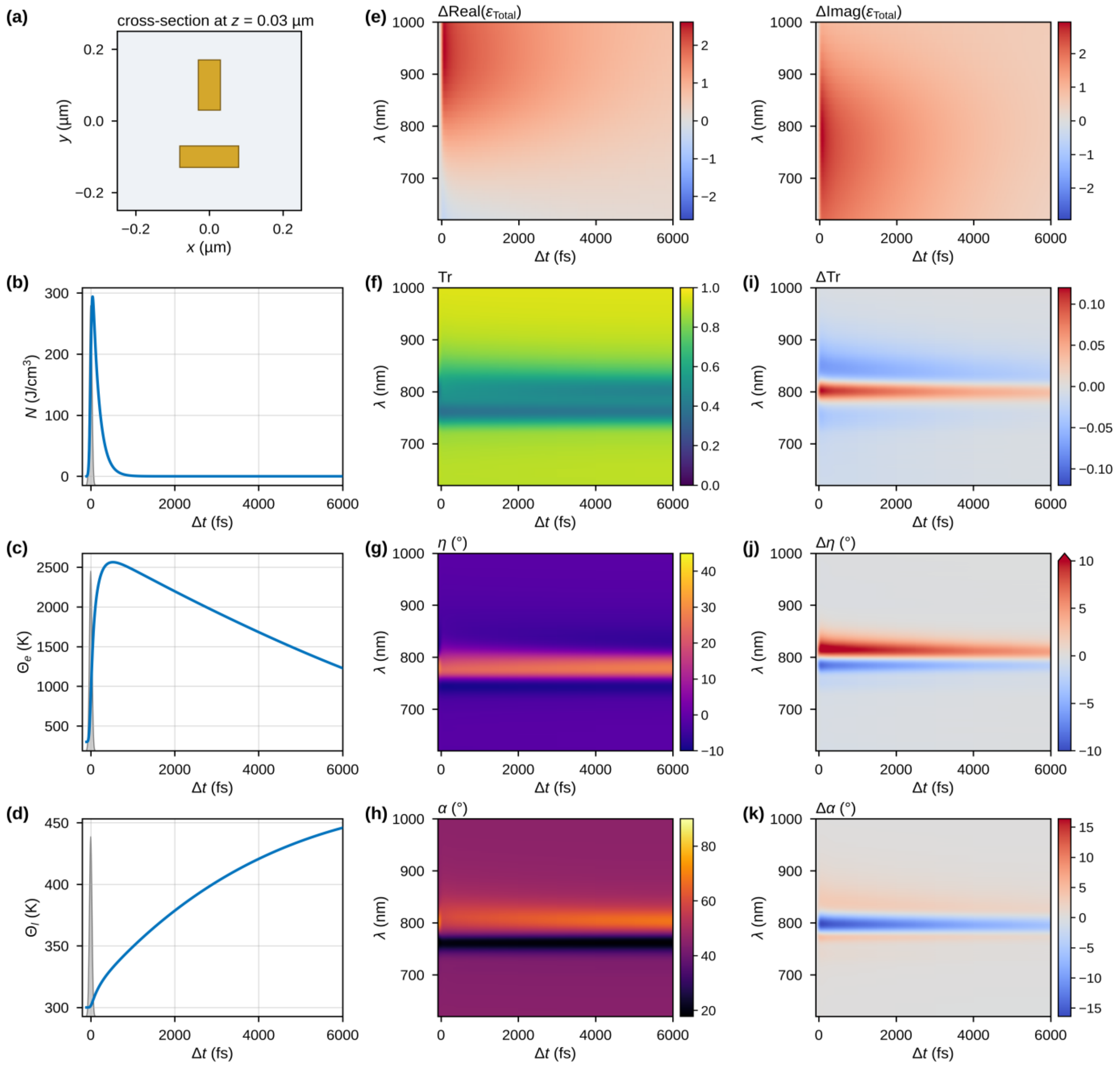

**Figure S6. Transient thermodynamic and optical response of the scaled orthogonal nanorod dimer metasurface on an n = 1.45 substrate. (a)** Cross-sectional schematic of the scaled unit cell geometry (z = 0.03 μm). **(b–d)** Thermodynamic response capturing the non-thermal electron energy density (*N*), electron temperature ($\Theta_e$), and lattice temperature ($\Theta_l$) under a ≈240 μJ/cm² pump. **(e)** Changes in the real and imaginary components of the Au dielectric function. **(f, i)** Absolute and differential transmission spectra mapping the transient bleaching and absorption. **(g, j)** Absolute ellipticity and transient modulation, reaching peak shifts of ≈11°. **(h, k)** Absolute optical rotation and corresponding ultrafast dynamic rotation reaching ≈16°.

## S6. Pump-induced phase changes of the two polarization channels

Figure S7 shows the pump-induced change of the transmission phase of the two polarization channels under the x-polarized pump of the main text (≈840 μJ/cm²). Because $\tau_{xy} = \tau_{yx} = 0$ and the probe is polarized at 45°, the phase changes of the transmitted $E_x$ and $E_y$ components equal $\Delta\mathrm{Arg}(\tau_{xx})$ and $\Delta\mathrm{Arg}(\tau_{yy})$. Only the phase of $\tau_{xx}$, whose resonance ($\mathrm{LSP_x}$) is pumped, changes appreciably. The y-oriented nanorod receives only ≈1% of the power absorbed from the x-polarized pump and is therefore kept at its equilibrium permittivity (Section S8 estimates the effect of this absorption); the small residual $\Delta\mathrm{Arg}(\tau_{yy})$ (below 1°) stems from the modified x-oriented nanorod. Together with the change of $|\tau_{xx}|$, this phase change determines the transient polarization state through Equation (5) of the main text.

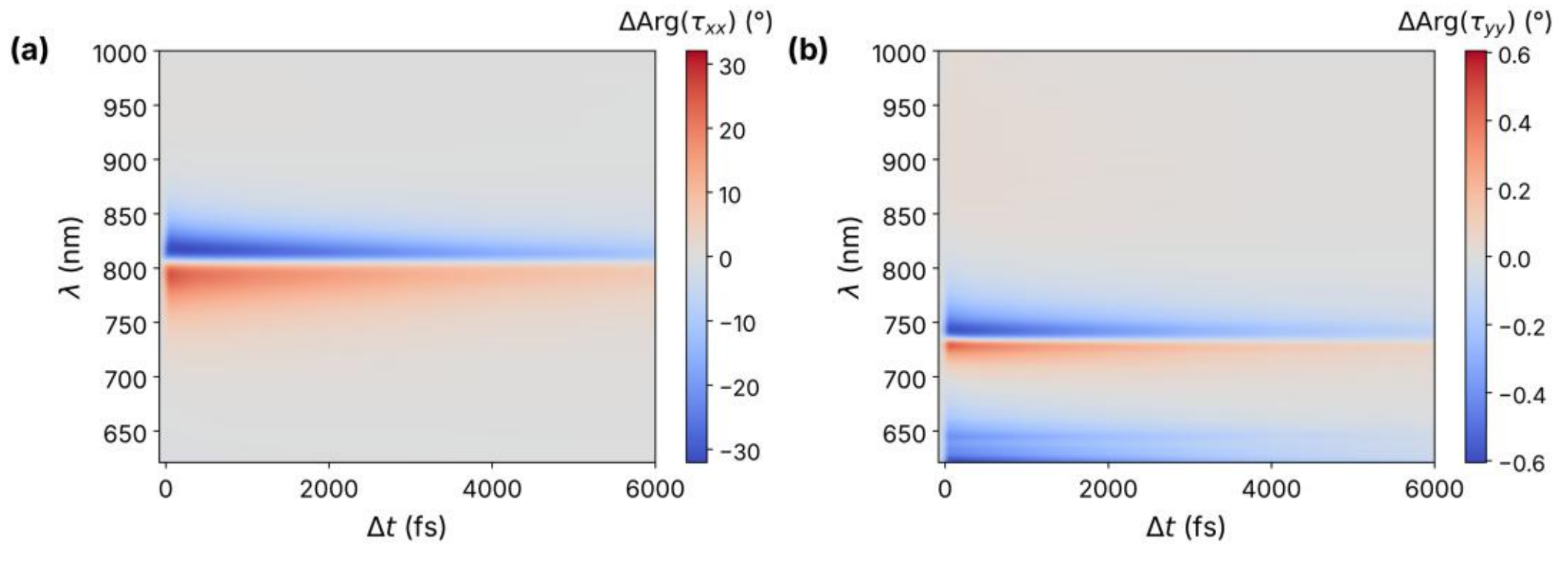

**Figure S7. Pump-induced change of the transmission phase of the two channels. (a)** $\Delta\mathrm{Arg}(\tau_{xx})$ and **(b)** $\Delta\mathrm{Arg}(\tau_{yy})$ versus probe wavelength and pump–probe delay (x-polarized pump, ≈840 μJ/cm²). Note the different colour scales: $\Delta\mathrm{Arg}(\tau_{xx})$ reaches −32° and +25°, whereas $\Delta\mathrm{Arg}(\tau_{yy})$ remains below 1°.

**S7. Response to other probe polarizations**

Because the Jones matrix of the unit cell remains diagonal under the x-polarized pump ($\tau_{xy}$ = $\tau_{yx}$ = 0, mirror symmetry $x \rightarrow -x$), a probe linearly polarized at an azimuth φ from the x axis is transmitted as $E \propto$ ($\tau_{xx}$ cos φ, $\tau_{yy}$ sin φ). The 45° probe of the main text therefore determines, at every delay, $\tau_{yy}/\tau_{xx}$ = ($s_2$ + i$s_3$)/(1 + $s_1$) from the normalized Stokes parameters of the transmitted light, and the channel transmittances $T_x$ = Tr(1 + $s_1$) and $T_y$ = Tr(1 − $s_1$), where Tr is the transmittance of the 45° probe. For any other azimuth, the polarization state follows from ($\tau_{yy}/\tau_{xx}$) tan φ and the transmittance from $T_x$ cos²φ + $T_y$ sin²φ, without further simulations. The phase of $\tau_{yy}/\tau_{xx}$ obtained in this way agrees with the separately stored phases of the two transmitted field components to within 0.002°. Figure 5 of the main text shows the resulting pump-induced changes at $\Delta t$ = 80 fs. It characterizes each change by its displacement on the Poincaré sphere, the angle between the normalized Stokes vectors with and without the pump, $d$ = arccos($\mathbf{s}\cdot\mathbf{s}_0$) with $\mathbf{s}$ = ($s_1$, $s_2$, $s_3$), and by its direction $\theta$ = arctan(|2Δ$\eta$|/|2Δ$\alpha$ cos 2$\eta_0$|), where $\eta_0$ is the unpumped ellipticity. Because a state has the longitude 2$\alpha$ and the latitude 2$\eta$ on the sphere, a rotation of the polarization ellipse moves it along a parallel ($\theta$ = 0°) and a change of ellipticity along a meridian ($\theta$ = 90°); for small changes, $d \approx [(2\Delta\eta)^2 + (2\Delta\alpha \cos 2\eta_0)^2]^{1/2}$. Direct simulations at every pump–probe delay confirm the reconstruction (Figure S8): with the probe polarized at φ = 27° and pump 1 alone, at the 22 delays of Figures 3 and 4 of the main text (−80 fs to 6 ps), and at φ = 37° with both pumps, at the 34 delays of Figure 6 of the main text. Over 700–900 nm, the transmittance agrees to within $3\cdot10^{-4}$, Δ$T$/$T$ to within 0.08%, and the pump-induced changes of $\eta$ and $\alpha$ and the displacement on the Poincaré sphere to within 0.05° ($\alpha$ where |$\eta$| ≤ 40° before and after the pump). For a probe polarized along x, at the largest change with pump 1 alone ($\Delta t$ = 80 fs) and right after pump 2 ($\Delta t$ = 580 fs), the cross-polarized fraction of the transmitted power stays below $2\cdot10^{-19}$ over the whole simulated spectrum (621–1000 nm; $|\tau_{yx}/\tau_{xx}| < 4\cdot10^{-10}$), and a non-zero $\tau_{xy}$ would spoil the agreement at 27° and 37°: the Jones matrix remains diagonal also when both nanorods are pumped (Section S8), because each nanorod keeps a uniform permittivity and hence the mirror symmetry. Figure 6e of the main text uses this for a probe at φ = 37°.

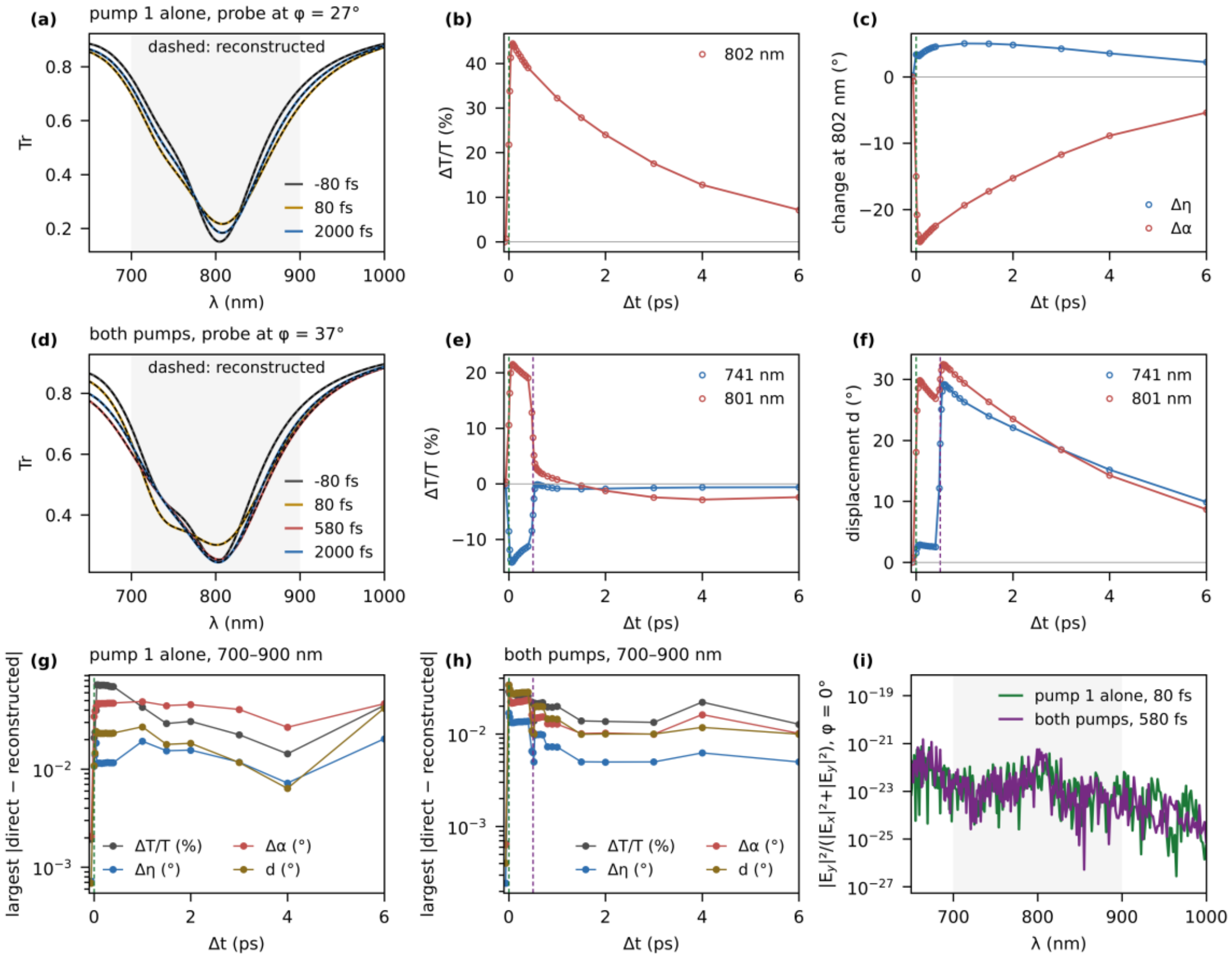

**Figure S8. Direct check of the reconstruction for other probe polarizations (Section S7), at every delay.** (a–c) Pump 1 alone (x-polarized, 800 nm, ≈840 µJ/cm²), probe linearly polarized at φ = 27° (setting D of Figure 5 of the main text): (a) transmittance before the pump ($\Delta t = -80$ fs), at 80 fs and at 2 ps; (b) $\Delta T/T$ and (c) $\Delta\eta$ and $\Delta\alpha$ at 802 nm versus delay. (d–f) Both pumps (Figure 6 of the main text), probe at φ = 37°: (d) transmittance at −80 fs, at 80 fs (pump 1 only), and at 580 fs and 2 ps (after pump 2); (e) $\Delta T/T$ and (f) displacement on the Poincaré sphere at 741 and 801 nm versus delay, as in Figure 6e of the main text. In (a) and (d), solid lines: direct simulation; dashed lines: reconstruction from the 45° probe. In (b), (c), (e) and (f), markers: direct simulation; lines: reconstruction; dashed vertical lines: arrival of pump 1 ($\Delta t = 0$) and of pump 2 ($\Delta t = 0.5$ ps). (g, h) Largest absolute difference between direct simulation and reconstruction over 700–900 nm at each delay: $\Delta T/T$ (in %), $\Delta\eta$, $\Delta\alpha$ (where $|\eta| \leq 40°$ before and after the pump) and the displacement $d$ (in degrees). (i) Cross-polarized fraction $|E_y|^2/(|E_x|^2 + |E_y|^2)$ of the transmitted light for a probe polarized along x (φ = 0°), which vanishes for a diagonal Jones matrix, with pump 1 alone (80 fs) and with both pumps (580 fs). Shaded: 700–900 nm.

## S8. Sequential excitation of both nanorods

In the simulation of Figure 6 of the main text, pump 1 (x-polarized, 800 nm, $LSP_x$) arrives at $\Delta t = 0$ and pump 2 (y-polarized, 740 nm, $LSP_y$) at $\Delta t = 500$ fs, both as 70-fs Gaussian pulses with an incident fluence of ≈840 μJ/cm². Each pump has its own absorption simulation and its own 3TM calculation, at the photon energy of its wavelength. The unit cell absorbs a fraction $A = 0.223$ of pump 1 and $A = 0.256$ of pump 2 (Figure S9d, e); treating each pump as monochromatic changes these fractions by less than ≈2%. The full absorbed fraction is assigned to the nanorod along which each pump is polarized, which actually receives 98.9% (pump 1) and 98.4% (pump 2) of it (see below). The electron temperature peaks at ≈2290 K in the x-oriented and at ≈2730 K in the smaller y-oriented nanorod (Figure S9c), and the lattice eventually warms by ≈150 and ≈210 K, far below the melting point of gold. As the two nanorods are electrically isolated and exchange negligible heat on these timescales, each takes at every probe delay the permittivity of its own 3TM calculation, at $\Delta t$ for the x-oriented and at $\Delta t - 500$ fs for the y-oriented nanorod (unexcited gold before its pump arrives). The 34 probe delays, from −80 fs to 6 ps, are dense around both pump arrivals. The response to pump 1 alone is taken from the simulation of Figures 3 and 4 of the main text, which has no delays between 400 fs and 1 ps; it is interpolated linearly there.

Each pump acts mainly near its own resonance. The largest displacement of the polarization state of the 45° probe on the Poincaré sphere (Section S7) produced by pump 1 before pump 2 arrives ($\Delta t \leq 400$ fs, before the leading edge of pump 2) peaks at ≈24° near 800 nm, whereas the additional displacement produced by pump 2, the angle between the states computed with and without it ($\Delta t = 500$–900 fs), peaks at ≈34° near 740 nm and stays below 5° beyond 810 nm. Only at ≈760–790 nm, where the two bands overlap, do the displacements produced by the two pumps both exceed 10°. The cancellation in Figure 6e of the main text tolerates small errors of the probe azimuth: after pump 2 ($\Delta t \geq 560$ fs), $|\Delta T/T|$ stays within 3% at 741 nm for $\varphi = 34$–40° (at $\varphi = 37°$, from 726 to 748 nm) and within 5% at 801 nm for $\varphi = 36$–41° (≈3% at 37°).

**Absorption of each pump in the other nanorod.** Flux monitors enclosing each nanorod, on a 2.5-nm mesh (absorbed power density in Figure S9a,b), show that 1.1% of the power absorbed from pump 1 is dissipated in the y-oriented nanorod and 1.6% of that from pump 2 in the x-oriented nanorod; these shares come largely from each nanorod's own absorption of light polarized across it. The simulations neglect them. To estimate their effect over 700–900 nm, the response of each nanorod, known from the simulations with pump 1 alone and with both pumps, was weighted with the energy that nanorod actually receives. With pump 1 alone, the transmittance then changes by less than 0.002 and the polarization state by at most ≈0.4–0.5°. For the sequence of Figure 6 of the main text, the polarization state changes by at most ≈0.4–0.5°, the change of Arg($\tau_{xx}$) caused by pump 2 rises from ≈0.7° to ≈1.0°, and the probe azimuth at which pump 2 best cancels the intensity change of pump 1 stays at 37°. The absorption of each pump in the other nanorod is therefore negligible.

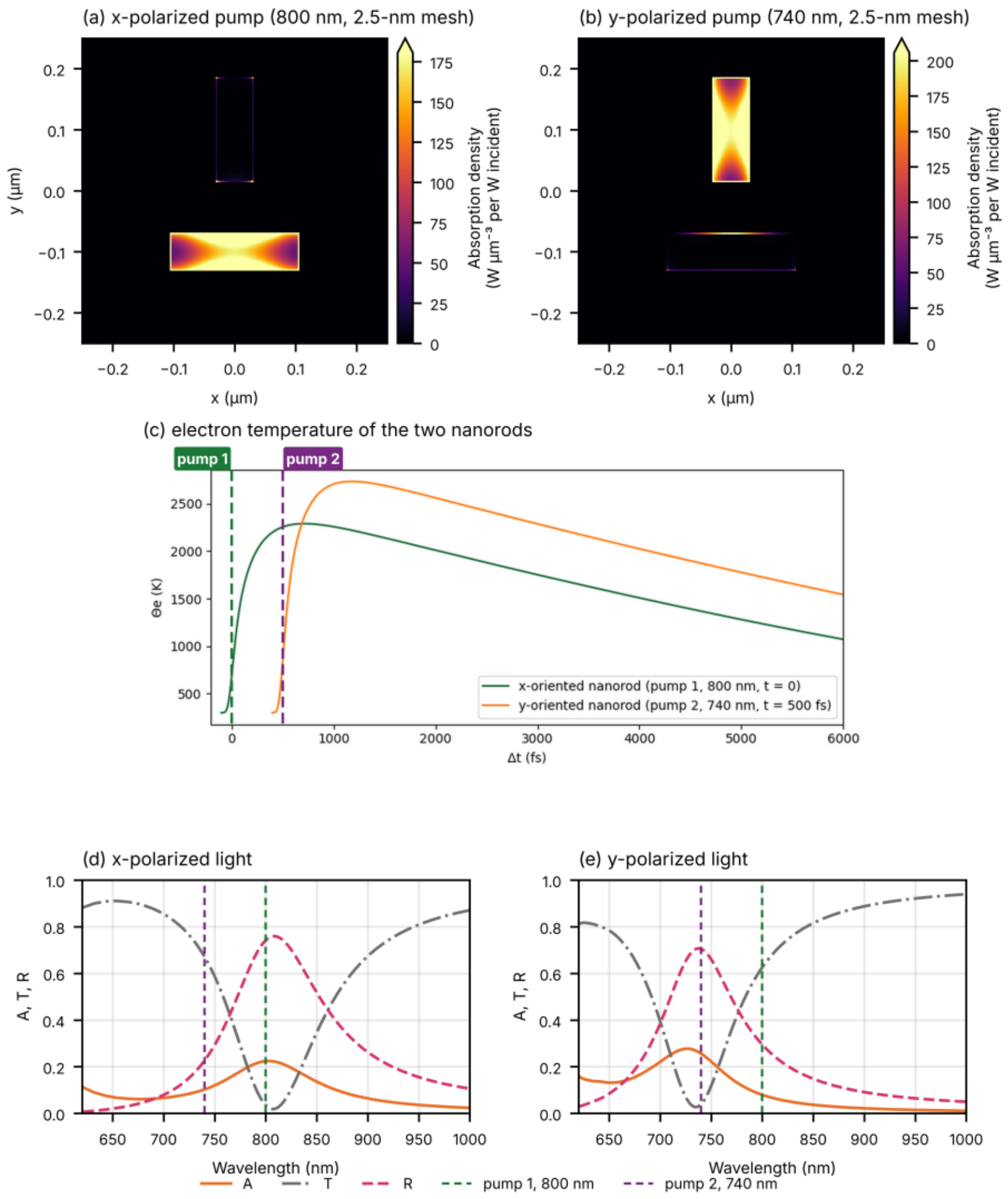

**Figure S9. Sequential excitation of the two nanorods (Figure 6 of the main text). (a, b)** Absorbed power density in the plane $z$ = 25 nm for **(a)** the x-polarized pump at 800 nm and **(b)** the y-polarized pump at 740 nm, computed on a 2.5-nm mesh; the colour scales are saturated to show weak absorption. The bright lines along the faces of the nanorods come from grid points on the metal surface, where the permittivity and the fields are interpolated, and overstate the absorption there. **(c)** Electron temperature $\Theta_e$ of the two nanorods from the two 3TM calculations; dashed

lines: arrival of pump 1 ($\Delta t = 0$) and pump 2 ($\Delta t = 500$ fs). **(d, e)** Absorptance A, transmittance T and reflectance R of the unpumped unit cell for **(d)** x- and **(e)** y-polarized light; dashed lines: the pump wavelengths, 800 nm (pump 1) and 740 nm (pump 2).